\newlength{\intwidth}

\documentclass[lineno]{jfm}

\usepackage{amssymb}

\usepackage{array}
\usepackage{amsmath}
\usepackage{latexsym}
\usepackage{bezier}
\usepackage{psfrag,graphicx}
\usepackage{bm}
\usepackage{float}
\usepackage{multirow}
\usepackage{mathrsfs}
\usepackage{color}
\usepackage{enumerate}
\usepackage{natbib}
\usepackage{nicefrac}
\usepackage[colorlinks, citecolor=blue]{hyperref}
\usepackage{overpic}
\usepackage{rotating}
\usepackage{verbatim}
\usepackage{enumitem}

\newcommand{\RealN}{\mathbb{R}}

\newtheorem{thm}{Theorem}

\begin{document}

\title[Analysis of inviscid shear instability of axisymmetric flows]{Analysis of inviscid shear instability of axisymmetric flows}

\author{
  Kengo Deguchi\aff{1}\corresp{\email{kengo.deguchi@monash.edu}},
  Haider Munawar\aff{1}  \and Runjie Song\aff{1}
}
 
 \affiliation{
   \aff{1}School of Mathematics, Monash University, VIC 3800, Australia
}

\maketitle

\begin{abstract}
Simple analytical criteria are derived to determine whether axisymmetric base flows in annuli and pipes are stable or unstable. Both axisymmetric and non-axisymmetric inviscid disturbances are considered. Our sufficient condition for stability improves upon the classical result of \cite{Batchelor_Gill_1962}, following the idea of the second Kelvin-Arnol'd stability theorem. A novel sufficient condition for instability is also derived by extending the recently proposed hurdle theorem for parallel flows \citep{Deguchi_Hirota_Dowling_2024}. These analytical criteria are applied to annular and pipe model flows and are shown to effectively predict the neutral parameters obtained from eigenvalue computations of the stability problem.
\end{abstract}

\section{Introduction}\label{sec:introduction}

Axisymmetric flows arise in a variety of applications, including flows through annuli and pipes as well as round jets, and their stability has long been of considerable interest. In most practical situations, the Reynolds number is high, so viscous effects may be neglected. 
The inviscid approximation greatly simplifies the stability equations and often leads to fruitful mathematical results. An important contribution to the inviscid stability of axisymmetric base flows was made by \cite{Batchelor_Gill_1962}, who derived a convenient sufficient condition for stability. This result can be viewed as an extension of the well-known Rayleigh-Fjortoft stability condition for parallel flows \citep{Rayleigh1880,Fjortoft1950}. Although significant progress has since been made in the inviscid stability analysis of parallel base flows, somewhat surprisingly, applications to axisymmetric base flows remain scarce.

\cite{Arnold1966} showed that two distinct sufficient conditions for stability exist through a pseudoenergy-based analysis, which is now known as Arnol'd's method (see \cite{Shepherd_1993} also). Earlier, Lord Kelvin had also suggested the presence of two such conditions \citep{Thomson1880}. Owing to this historical development, those stability conditions are called the Kelvin-Arnol'd $1^{\rm st}$ and $2^{\rm nd}$ shear-stability theorems (KA-I and KA-II).  
The Rayleigh-Fj{\o}rtoft condition corresponds to KA-I, as does the result of \cite{Batchelor_Gill_1962}. For parallel flows, the KA-II conditions have also been well studied, particularly in the context of instabilities in planetary atmospheres \cite[see][for example]{Stamp_Dowling_1993,Dowling2020,RD26}. Therefore, one might expect that the KA-II condition could also be applied to axisymmetric base flows to yield useful stability criteria; however, no such results have been explicitly derived in the literature.

Except for a few special cases, the KA stability conditions generally fail to sharply detect points of neutral stability, as first demonstrated in the pioneering work of \cite{Tollmien1935}. This limitation has motivated efforts to identify classes of base flows that admit sharp stability criteria and to simplify those criteria as much as possible. For example, \cite{Rosenbluth_Simon1964} and \cite{Balmforth_Morrison1999} employed the Nyquist method, whereas \cite{Barston1991} and \cite{Hirota_etal2014} drew inspiration from the properties of certain quadratic forms. Despite being more numerically tractable than the full eigenvalue problem, these approaches still rely on the solution to a Fredholm integral equation or similarly complex criteria.

Recently, \cite{Deguchi_Hirota_Dowling_2024} derived a simple sufficient condition for instability. The condition essentially depends on whether the reciprocal Rossby-Mach number, derived from the base flow, surpasses a threshold called the \textit{hurdle}. Since this can be conveniently assessed using a simple graphical procedure, the result is referred to as the \textit{hurdle theorem}. Although this condition does not sharply determine the neutral point, plotting the regions in parameter space where the hurdle theorem and the KA criteria hold allows one to roughly estimate the location of neutral points. In practice, this helps restrict the parameter range that needs to be explored in the eigenvalue problem.

The proof by \cite{Deguchi_Hirota_Dowling_2024} follows the two-step approach that became popular after the seminal work of \cite{Howard1964}: (i) first, demonstrate the existence of a neutral solution, and (ii) then show that an unstable mode emerges when this neutral state is perturbed. The essence of this approach can already be found in \cite{Tollmien1935}. A mathematically rigorous proof of step (ii) was later provided by \cite{Lin2003}, and more recently simplified by \cite{Kumar2025}. In step (i), the neutral solution is constructed through minimisation of the Rayleigh quotient. The central idea of \cite{Deguchi_Hirota_Dowling_2024} is to analytically derive a simple condition ensuring the existence of a neutral solution by estimating the minimum value with a suitable trial function.

In this paper, we extend the sufficient stability condition of \cite{Batchelor_Gill_1962} and derive a simple sufficient condition for instability for axisymmetric base flows. The former is obtained following the aforementioned KA-II framework, while the latter is derived by generalising the hurdle theorem. For annular flows, the analysis proceeds in much the same way as for parallel flows, although the azimuthal wavenumber of the perturbations must be treated carefully. Pipe flows, however, require additional consideration, and the conditions must be modified from those for the annular case to obtain satisfactory results. 

The proposed stability conditions are applied to representative model flows to assess their practical usefulness. More specifically, we consider cases where shear flow and thermal convection coexist. Parallel shear flows, whether driven by a pressure gradient or by moving boundaries, tend to be inviscidly stable, as exemplified by plane Couette flow or plane Poiseuille flow. The first model flow we consider is sliding Couette flow (an annular version of Couette flow, see \cite{Gittler_1993}, \cite{DEGUCHI_NAGATA_2011}), coupled with thermal convection due to the temperature difference between the walls. The flow is designed so that, 
in the absence of the wall sliding effect, the system reduces to natural convection in a vertical annulus, a configuration frequently studied in the natural convection literature (\cite{Choi_Korpela_1980},\cite{Yao_Rogers_1989}, \cite{Kang_Yang_Mutabazi_2015},\cite{Wang_Chen_2022}). The second model is a vertically oriented Hagen-Poiseuille flow with homogeneous internal heating. This model can be regarded as the pipe analogue of the channel model flow used to test the hurdle theorem in \cite{Deguchi_Hirota_Dowling_2024}, and it was also investigated by \cite{Senoo2012} and  \cite{Marensi_He_Willis_2021}.
The performance of the stability conditions will be assessed by comparing their predictions with the numerically obtained  neutral points.
Those inviscid stability results are further compared with linearised Navier–Stokes computations. 

The structure of the paper is as follows. In the next section, we formulate the stability problem for axisymmetric base flows. Numerical methods for both inviscid and viscous stability problems are also introduced. Section~\ref{sec:3} summarises the sufficient conditions for stability and instability derived in this paper. These conditions are applied to model flows in annuli and pipes in section~\ref{sec:4}, with their derivation presented in section~\ref{sec:5}. Finally, section~\ref{sec:6} presents the conclusions. In the same section, we also discuss the implications of the present results for round jets, which originally motivated the work of \cite{Batchelor_Gill_1962}.

\section{Formulation of the problem}\label{sec:2}

\subsection{Inviscid stability problem for axisymmetric base flows}\label{sec:2.1}
Let $\mathbf{u}=u_r\mathbf{e}_r+u_{\varphi}\mathbf{e}_{\varphi}+u_z\mathbf{e}_z$ and $p$ denote the velocity and pressure of an incompressible fluid in cylindrical coordinates $(r,\varphi,z)$, with $\mathbf{e}_r$, $\mathbf{e}_{\varphi}$, and $\mathbf{e}_z$ the associated unit vectors. 
Periodicity of $2\pi/k$ is imposed in the $z$ direction, with $k$ denoting the axial wavenumber.
We assume a steady axisymmetric base flow $\mathbf{u}=U(r)\mathbf{e}_z$ directed along the axial direction, and assess its stability by examining the evolution of infinitesimal perturbations $\tilde{\mathbf{u}}=\tilde{u}_r\mathbf{e}_r+\tilde{u}_{\varphi}\mathbf{e}_{\varphi}+\tilde{u}_z\mathbf{e}_z$ superposed on it. The azimuthal wavenumber of the perturbation is denoted by $n$. The physical requirement of $2\pi$ periodicity in the $\varphi$ direction implies that $n$ must be an integer.

The stability of the flow can be analysed by solving the linearised Navier-Stokes equations; if viscosity is neglected, the analysis can begin with the linearised Euler equations.
\cite{Batchelor_Gill_1962} showed that the latter \textit{inviscid stability problem} boils down to solving the differential equation 
\begin{eqnarray}\label{Geq}
\{\frac{r}{N^2+r^2}(rG)' \}'-k^2G+\frac{Q'}{U-c}rG=0,\qquad Q=\frac{-rU'}{N^2+ r^2}
\end{eqnarray}
with suitable boundary conditions. 
Here, a prime 
denotes ordinary differentiation with respect to radius, $r$. For a fixed wavenumber ratio $N = n/k$, equation (\ref{Geq}) together with suitable boundary conditions constitutes an eigenvalue problem, with the complex phase speed $c=c_r+ic_i$ as the eigenvalue. 
By symmetry, it suffices to consider $k > 0$ and $n \geq 0$. From the eigenfunction $G(r)$, the perturbation field can be reconstructed as
\begin{eqnarray}
~[\tilde{u}_r,\tilde{u}_{\varphi},\tilde{u}_{z},\tilde{p}]=[iG(r),H(r),F(r),P(r)]e^{in\varphi+ik(z-ct)}+\text{c.c.},
\end{eqnarray}
where
\begin{eqnarray}
\left [
\begin{array}{c}
F\\
H
\end{array}
\right ]
=
\frac{1}{k(N^2+r^2)}
\left [
\begin{array}{cc}
-N & -r\\
r & -N
\end{array}
\right ]
\left [
\begin{array}{c}
NU'G/(U-c)\\
rG'+G
\end{array}
\right ],\\
P=-(U-c)F-k^{-1}U'G,
\end{eqnarray}
and c.c. denotes the complex conjugate. 
The existence of an eigenvalue with $c_i > 0$ indicates the presence of disturbances that grow exponentially 
with time $t$, and hence the flow is unstable. The flow is said to be stable when no such unstable mode exists (the mode with $c_i=0$ is neutrally stable).

The general properties of the inviscid stability problem (\ref{Geq}) are the main focus of this paper. For concreteness, however, we examine the stability conditions for specific base flows. The model base flows to be introduced in section~\ref{sec:4} can be described using the Navier-Stokes equations together with the Boussinesq approximation. We specifically consider the case of an infinitesimally small Prandtl number, for which the stability problem reduces to the linearised Navier-Stokes equations. 
Note that, since the Prandtl number is the ratio of viscous to thermal diffusion, a small value implies that the diffusion term dominates advection in the temperature equation. Consequently, temperature perturbations decay and the buoyancy term is removed from the momentum equations for perturbations (see \cite{L_1999} for example).

Section~\ref{sec:4.1} examines a flow between coaxial cylinders, while section~\ref{sec:4.2} studies a pipe flow.
For the annular geometry, we set the inner and outer cylinder walls at $r=r_i$ and $r=r_o$. The radius ratio $\eta=r_i/r_o \leq 1$ is the only parameter needed to specify the geometry. Under the nondimensionalisation in which the gap is set to 2, 
\begin{eqnarray}
r_i=2\eta/(1-\eta), \qquad r_o=2/(1-\eta).
\end{eqnarray}
The no-penetration boundary condition then requires that 
$G$ vanishes at the walls.

For the pipe flow problem, physical quantities are required to be \textit{regular} at the centreline. If we assume that the perturbation velocity field is Taylor expandable about the pipe axis in Cartesian coordinates, then $G$ has the following expansion (see \cite{Eisen_1991}):
\begin{subequations}\label{taylor}
\begin{eqnarray}
G=\mathcal{G}_1r+\mathcal{G}_3r^3+\cdots, ~~~~&&\text{if}~~~~ n= 0,\\
G=\mathcal{G}_{n-1}r^{n-1}+\mathcal{G}_{n+1}r^{n+1}+\cdots ~~~~&&\text{if}~~~~ n\neq 0.
\end{eqnarray}
\end{subequations}
Therefore, the eigenfunction for the pipe problem is expected to satisfy
\begin{subequations}\label{pipecentre}
\begin{eqnarray}
G'=0, ~~~~\text{if}~~~~ n= 1,\\
G=0 ~~~~\text{if}~~~~ n\neq 1,
\end{eqnarray}
\end{subequations}
at $r=0$. 
Note that (\ref{pipecentre}) is not a boundary condition of (\ref{Geq}).
This conclusion follows from the standard singularity theory of ordinary differential equations. As commented in \cite{Lessen_Singh_1973}, 
the method of Frobenius expansion yields that one of the linearly independent solutions is singular, while the other, non-singular solution satisfies (\ref{pipecentre}).

\subsection{Numerical method}\label{sec:2.2}

The eigenvalue problem (\ref{Geq}) can be solved using the Chebyshev-collocation method, in which $G(r)$ is expanded by the basis functions $\Psi_l(r)$:
\begin{eqnarray}
G(r)=\sum_{l=0}^L \hat{G}_l \Psi_l(r).
\end{eqnarray}
The annular case is straightforward. We use the basis $\Psi_l(r)=(1-y^2)T_l(y)$, where $T_l(y)$ are the Chebyshev polynomials in the shifted variable $y=r-r_m $. 
To map $y$ into the interval $(-1,1)$, we set $r_m=(r_o+r_i)/2$, the mid-gap. Collocation points are chosen as $r_j=r_m+\cos(\frac{j+1}{L+2}\pi)$,  $j=0,1,\dots,L$.

For the pipe case, we checked the consistency of the numerical
codes using the following two methods:
\begin{enumerate}[label=(\Alph*)]
\setlength{\itemsep}{0.2cm}

\item ~Expand $G$ in modified Chebyshev polynomials
(see \cite{Deguchi_Walton_2013}). 
\begin{eqnarray}
\Psi_l(r)=
\left \{
\begin{array}{c}
(1-r^2)T_{2l+1}(r),\qquad \text{if $n$ is even},\qquad  \\
(1-r^2)T_{2l}(r),\qquad \text{if $n$ is odd}.
\end{array}
\right .
\end{eqnarray}
The governing equation is then evaluated at the collocation points $r_j=\cos((j+1)\pi/(2L+4)), j=0,1,\dots,L$.

\item ~Treat the domain as an annulus with a virtual inner cylinder of small radius $\epsilon>0$, and impose condition (\ref{pipecentre}) there. 
The code developed in \cite{Song_Dong_2023A} is used, which employs the fourth-order compact finite-difference scheme \cite[see][]{Malik1990}. 
We choose $\epsilon=0.001$.

\end{enumerate}

~\\
The results of the inviscid analysis are compared with the calculations of the following  linearised Navier–Stokes equations:
\begin{subequations}\label{linNS}
\begin{eqnarray}
k(U-c)G=P'-\frac{i}{Re}(G''+\frac{G'}{r}-(k^2+\frac{n^2+1}{r^2})G-\frac{2n}{r^2}H),\\
k(U-c)H=-\frac{n}{r}P-\frac{i}{Re}(H''+\frac{H'}{r}-(k^2+\frac{n^2+1}{r^2})H-\frac{2n}{r^2}G),\\
k(U-c)F+U'G=-kP-\frac{i}{Re}(F''+\frac{F'}{r}-(k^2+\frac{n^2}{r^2})F),\\
G'+\frac{G}{r}+\frac{n}{r}H+kF=0,
\end{eqnarray}
\end{subequations}
where $Re$ is the Reynolds number. To solve the above eigenvalue problem, we employed a numerical code using the Chebyshev collocation method originally developed by \cite{DEGUCHI_NAGATA_2011} for the annular domain. For the pipe flow problem, the basis functions are replaced with those of \cite{Deguchi_Walton_2013}. The codes have been thoroughly tested and validated against results from other research groups (see, e.g., \cite{Deguchi_2017_PRE}; \cite{He_Deguchi_Song_Blackburn_2025}).

\section{Stability conditions} \label{sec:3}

Throughout the paper, we assume that $U(r)$ possesses $C^1$ smoothness in $\Omega$ (i.e., the first derivative is continuous in the domain). 
Here, for the annular case, $\Omega=(r_i,r_o)$, and for the pipe case, $\Omega=(0,1)$. For the pipe problem, we further assume that the even extension of $U(r)$ is $C^1$ in $(-1,1)$.

To state the main result of this paper, it is convenient to define
\begin{eqnarray}\label{defW}
W_{\alpha,N}(r)=\frac{rQ'}{U-\alpha}.
\end{eqnarray}
\cite{Batchelor_Gill_1962} also used a similar quantity in their analysis; however, note that our $Q$ is defined differently from theirs. 
The right hand side of (\ref{defW}) depends on a constant $\alpha$ and the wavenumber ratio $N$ via $Q$, hence the subscripts.
%

The main results of this paper are presented in the following theorems, which provide convenient criteria for assessing the inviscid stability problem (\ref{Geq}). 

\begin{thm}\label{KA}
The base flow is stable if the KA-I and/or KA-II conditions are satisfied for all $N$.
\begin{itemize}
\item KA-I: There exists $\alpha \in \RealN$ such that $W_{\alpha,N}\leq 0$ for all $r \in \Omega$. 
\item KA-II: There exists $\beta \in \RealN$ such that $W_{\beta,N}=\frac{rQ'}{U-\beta}\in [0,H(r)]$ for all $r \in \Omega$.
Here
\begin{eqnarray}\label{HNHN}
H=\left \{
\begin{array}{c}
(\frac{\pi}{\ln \eta})^2 \frac{1}{N^2+r_o^2}+\frac{1}{N^2}\qquad \text{if} \qquad N\geq 1,\\
(\frac{\pi}{\ln \eta})^2 \frac{1}{r_o^2}\qquad \text{if} \qquad N= 0
\end{array}
\right .
\end{eqnarray}
for the annular case,
\begin{eqnarray}\label{HNHN2}
H=\left \{
\begin{array}{c}
 (j_{1,1}^2r^2+N^{-2})/(N^2+1)\qquad \text{if} \qquad N\geq 1,\\
j_{1,1}^2 \qquad \text{if} \qquad N= 0
\end{array}
\right .
\end{eqnarray}
for the pipe case ($j_{1,1}\approx 3.8317$ is the first positive zero of the Bessel function of the first kind of order one).
\end{itemize}
\end{thm}

\begin{thm}\label{Hannulus}
The base flow is unstable if the following conditions are satisfied for some $N$:
\begin{itemize}
\item $Q'(r)$ has only one zero at $r=r_c$ and it is isolated.
\item $Q''(r_c)\neq 0$.
\item $U'(r_c)\neq 0$.
\item $W_{\alpha,N}>h(r)$ with  $\alpha=U(r_c)$ for all $r\in \Omega$. 
Here
\begin{eqnarray}\label{hthm2}
h=\frac{(\pi/\ln\eta)^2}{N^2+ r_i^2}+\frac{1}{N^2}
\end{eqnarray}
for the annular case,
\begin{eqnarray}\label{hthm3}
h=Cr^p, \qquad p\geq 2, \qquad C=\frac{1}{\rho_p}(\rho_N+\frac{1}{6N^2}),\nonumber \\
\rho_N=\frac{(3N^2+1)^2}{2}\ln (\frac{N^2+1}{N^2})-\frac{3(6N^2+1)}{4},\nonumber\\
\rho_p=\frac{8}{(p+2)(p+4)(p+6)}
\end{eqnarray}
for the pipe case. 
\end{itemize}
\end{thm}

Theorem \ref{KA}  corresponds to KA-I and KA-II for axisymmetric base flows, and guarantees regions in parameter space where the base flow is stable. KA-I is due to \cite{Batchelor_Gill_1962}, whereas KA-II is a new result.
Note that when $Q'$ has no zeros for any $N$, the base flow is stable, since in this case one can choose $|\alpha|$ sufficiently large to fulfill the KA-I condition. This is an extension of Rayleigh’s inflection point theorem. 

Theorem \ref{Hannulus} serves as a handy means of demonstrating the existence of instability in the parameter space. It extends the hurdle theorem by \cite{Deguchi_Hirota_Dowling_2024}; instability emerges when $W_{\alpha,N}$ surpasses the hurdle $h$.
\cite{Deguchi_Hirota_Dowling_2024} defined the  `reciprocal Rossby-Mach number' so that the hurdle height equals unity, inspired by studies of Jupiter’s jets.
An analogous quantity can be defined by multiplying $W_{\alpha,N}$ by a constant factor. Readers interested in the interpretation of our stability conditions using that number are referred to Appendix \ref{sec:AppA4}.

The proofs of above theorems are given in section~\ref{sec:5}.

\section{Analysis of model profiles} \label{sec:4}

\subsection{Flow through an annulus} \label{sec:4.1}

\begin{figure}
\centering
\begin{overpic}[width=0.5 \textwidth]{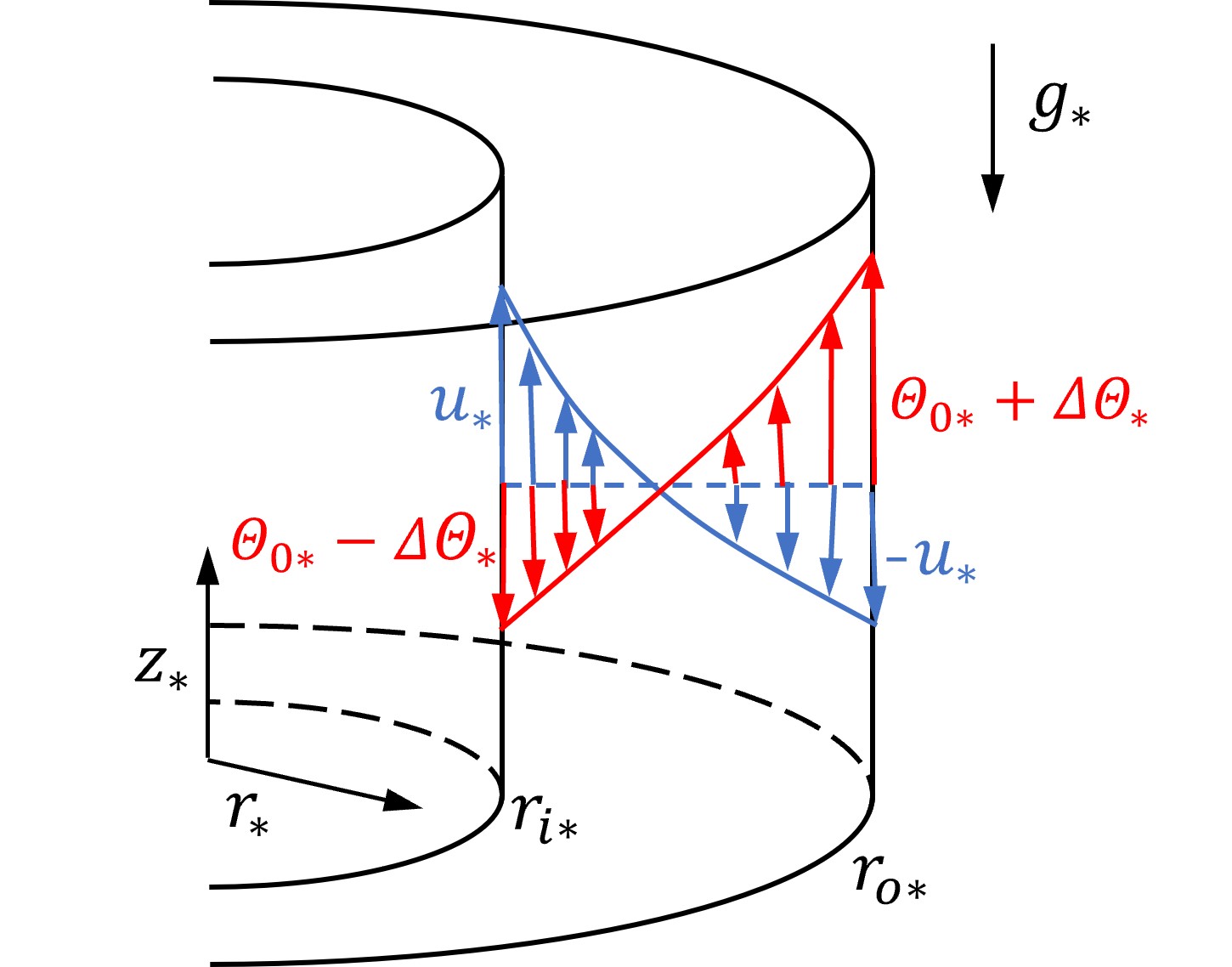}
\put(0,73){(a)}
\end{overpic}
\begin{overpic}[width=0.45 \textwidth]{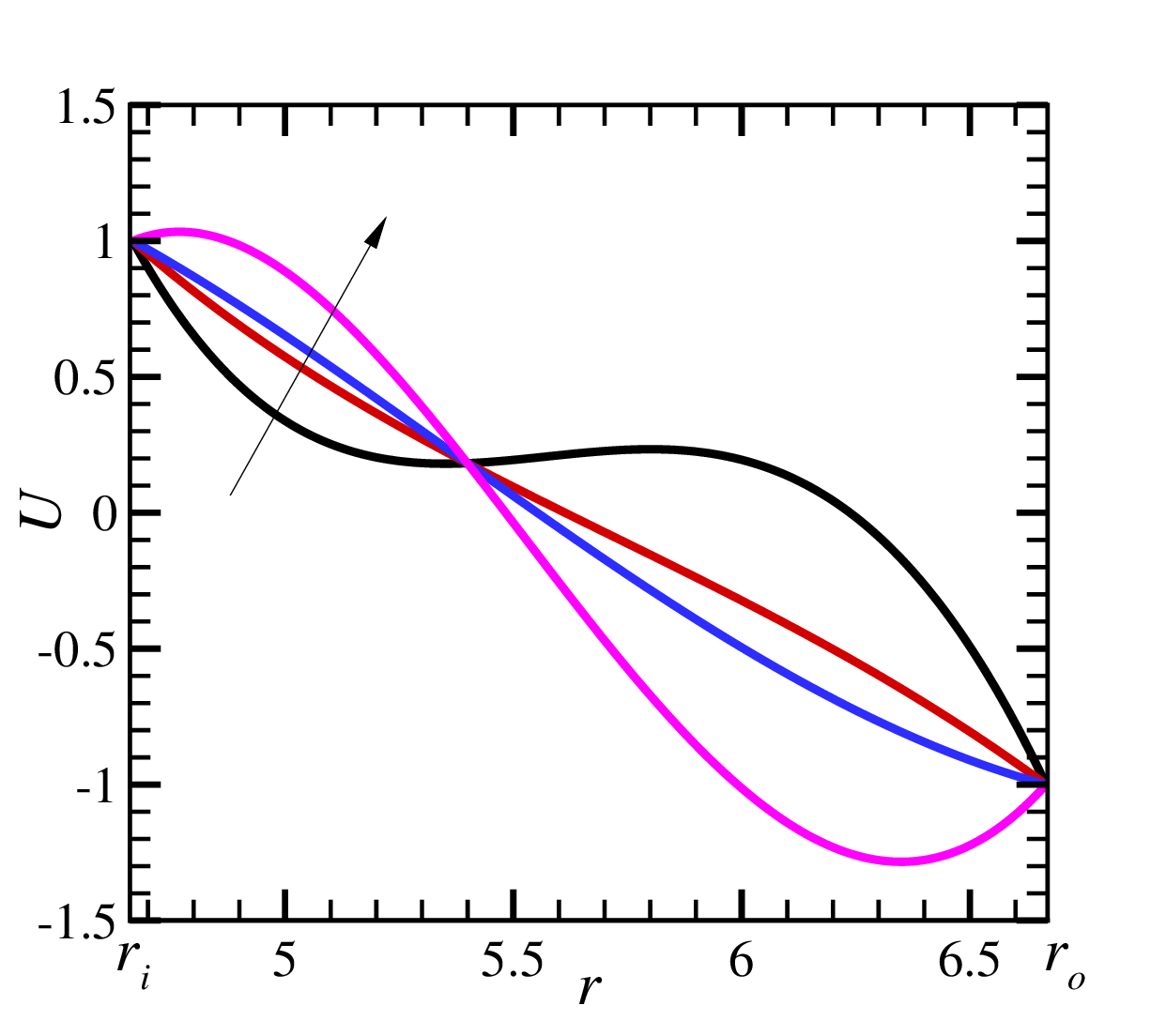}
\put(-3,80){(b)}
\put(35,70){Increasing $\chi$}
\end{overpic}
\caption{The annular model flow used in section~\ref{sec:4.1}. 
(a) Schematic of the model flow in the dimensional cylindrical coordinates $(r_*,\varphi,z_*)$.
Here, the radii $r_{i*}$ and $r_{o*}$ satisfy $r_{o*}-r_{i*}=2d_*$, $r_{i*}/r_{o*}=\eta$. (b) The base flow $U(r)$ given in (\ref{AnnularU}) for $\chi=-7,-1,1,7$. }
\label{fig:fig1}
\end{figure}

Consider the flow of a fluid between co-axial cylinders whose common axis is aligned with gravity (see figure~\ref{fig:fig1}-(a)). The fluid motion is driven by two effects: the differential axial movement of the cylinders and buoyancy resulting from an imposed temperature difference between the cylinders. 

We take half of the cylinder gap $d_*$, half of the differential speed $u_*$, and half of the imposed temperature difference $\Delta \theta_*$ as the characteristic length, velocity, and temperature scales, respectively. 
The non-dimensional Navier–Stokes equations under the Boussinesq approximation yield the governing equations for the axial base velocity $U(r)$ and the base temperature $\Theta(r)$:
\begin{subequations}\label{Baseeq1}
\begin{eqnarray}
\Theta''+r^{-1}\Theta'=0,\\
U''+r^{-1}U'=\chi \Theta.
\end{eqnarray}
\end{subequations}
The left hand sides represent diffusivity, while the term proportional to $\chi$ corresponds to buoyancy.
The ratio of the Grashof number $Gr$ to the Reynolds number $Re$ defines the parameter $\chi$ as follows.
\begin{eqnarray}
\chi=\frac{Gr}{Re},\qquad
Gr=\frac{\gamma_*g_* d_*^3 \Delta \theta_*}{\nu_*^2},\qquad Re=\frac{u_*h_*}{\nu_*}.
\end{eqnarray}
Here, $g_*$ denotes the gravitational acceleration,  $\gamma_*$ the thermal expansion coefficient, and $\nu_*$ the kinematic viscosity of the fluid.
The boundary conditions are
\begin{subequations}\label{BC1}
\begin{eqnarray}
U=1, ~~\Theta=-1\qquad \text{at}\qquad r=r_i,\\
U=-1, ~~\Theta=1\qquad \text{at}\qquad r=r_o.
\end{eqnarray}
\end{subequations}
The limit $\chi \rightarrow 0$ corresponds to sliding Couette flow (\cite{Gittler_1993},\cite{DEGUCHI_NAGATA_2011}), while
for large $|\chi|$, the system approaches natural convection in a vertical annulus (\cite{Choi_Korpela_1980},\cite{Yao_Rogers_1989}, \cite{Kang_Yang_Mutabazi_2015},\cite{Wang_Chen_2022}).

The solution of (\ref{Baseeq1}) with the boundary conditions (\ref{BC1}) is given by
\begin{eqnarray}
\Theta &=& C_1 \ln r+C_2,\\
U&=&C_3\ln r+C_4 
+\frac{\chi}{4} r^2(C_1-C_2- C_1 \ln r ), \label{AnnularU}
\end{eqnarray}
 where
\begin{eqnarray*}
C_1=-\frac{2}{\ln \eta},~~C_2=-1- C_1\ln r_i,~~C_3=\frac{C_5-C_6}{\ln \eta},~~C_4=C_5-C_3\ln r_i,\\
C_5=1+ \chi\frac{r_i^2}{4}\{ C_1\ln r_i-C_1+C_2 \},\qquad C_6=-1+\chi\frac{r_o^2}{4}\{ C_1 \ln r_o-C_1+C_2 \}.
\end{eqnarray*}
The base flow depends on $\eta$ and $\chi$. Figure~\ref{fig:fig1}-(b) shows the profiles at $\eta=0.7$ for various values of $\chi$.

The function $Q$ defined in (\ref{Geq}) is readily found. Its derivative, required for computation of $W_{\alpha,N}$ in (\ref{defW}), is given by
\begin{eqnarray}\label{Q1annular}
Q'=r\frac{N^2\chi  (C_2+C_1\ln r)+2C_3+\chi r^2 \frac{C_1}{2}}{(N^2+r^2)^2}.
\end{eqnarray}
If $N$ is sufficiently large, $Q'$ varies monotonically and therefore has at most one zero. This is convenient for applying Theorem \ref{Hannulus}.

\begin{figure}
\centering
\begin{overpic}[width=0.7 \textwidth]{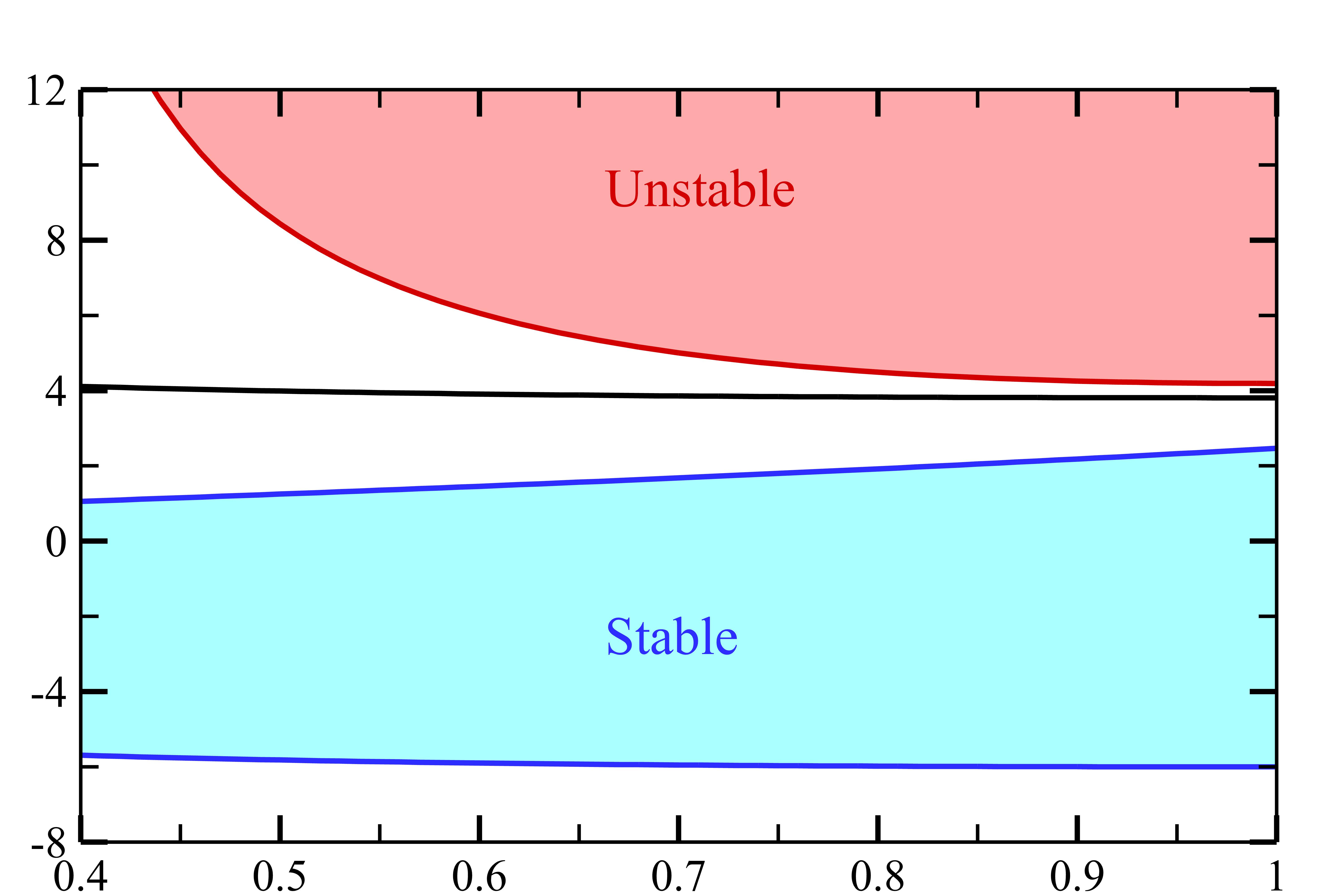}
\put(-1,32){\begin{turn}{90}{$\chi$}\end{turn}}
\put(50,-2.5){{$\eta$}}
\end{overpic}
\caption{Stability diagram of the annular model flow (figure \ref{fig:fig1}-(a)) in the $\chi$--$\eta$ plane. All physically possible wavenumbers are covered.
The black solid line represents the neutral curve of the inviscid problem (\ref{Geq}). Stability is guaranteed by Theorem \ref{KA} in the blue region, while unstable modes are found by Theorem \ref{Hannulus} in the red region.}
\label{fig:fig2}
\end{figure}
Figure~\ref{fig:fig2} summarises the inviscid stability in the $\eta$–$\chi$ parameter plane. The blue region indicates stability guaranteed by Theorem \ref{KA}, while the red region indicates instability guaranteed by Theorem \ref{Hannulus}. As expected, the neutral curve obtained by numerically solving the differential equation (the solid black line) lies between the two regions.

\begin{figure}
\centering
\begin{overpic}[width=0.8 \textwidth]{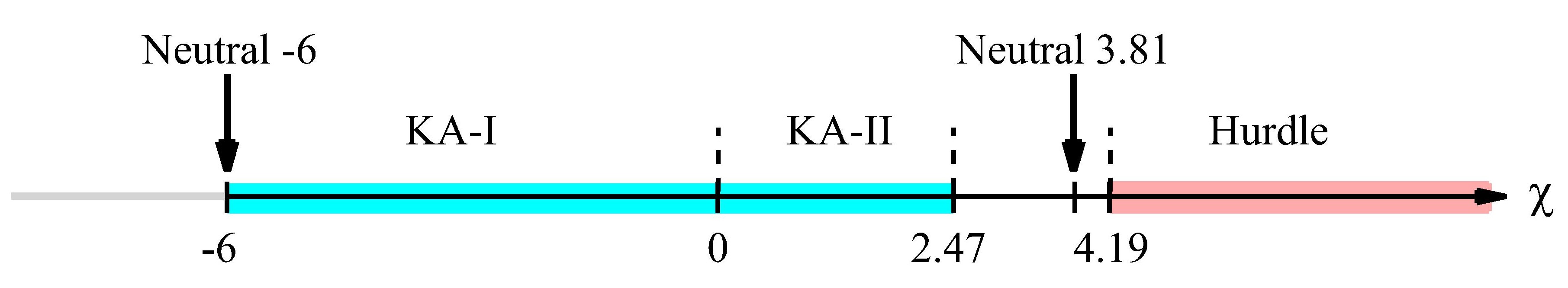}
\end{overpic}
\caption{Stability diagram of the annular model flow 
at the narrow-gap limit $\eta\rightarrow 1$. 
The eigenvalue problem (\ref{GeqNG}) indicates the presence of unstable modes for $\chi <-6$ and $\chi>3.81$.
The grey line shows that the profile of $W_{\alpha,N}$, given in (\ref{narrowW}), becomes singular when $\chi\leq -6$.
}
\label{fig:fig3}
\end{figure}

In the limit $\eta\rightarrow 1$, the analysis becomes particularly simple, so we begin our discussion from this case. 
This limit, referred to as the narrow-gap limit, corresponds to $r_i,r_o\rightarrow\infty$ in our non-dimensionalisation where the gap width is set to 2.
We introduce the new variable $y=r-r_m \in (-1,1)$, where $r_m=(r_o+r_i)/2$ is the mid-gap. 
Taking the limit $N\rightarrow \infty$ and $\eta\rightarrow 1$ while keeping $N\ll r_m$, we obtain
\begin{eqnarray}
W_{\alpha,N}=-\frac{1}{U-\alpha}\frac{d^2U}{dy^2},\qquad H=h=(\frac{\pi}{2})^2,
\end{eqnarray}
from (\ref{defW}), (\ref{HNHN}) and (\ref{hthm2}). Theorem \ref{Hannulus} reduces to the hurdle theorem for parallel flows derived in \cite{Deguchi_Hirota_Dowling_2024}.
The limiting form of the base flow is 
\begin{eqnarray}
U=\frac{\chi}{6}y(y^2-1)-y.
\end{eqnarray}
The numerator of $W_{\alpha,N}$, i.e. $\frac{d^2U}{dy^2}$, 
vanishes only at $y=y_c=0$. Setting $\alpha=U(y_c)=0$, we have
\begin{eqnarray}\label{narrowW}
W_{\alpha,N}=\frac{6\chi}{6+\chi (1-y^2)}.
\end{eqnarray}

The results of applying Theorems \ref{KA} and \ref{Hannulus} for the narrow-gap limit case, shown in figure~\ref{fig:fig3}, can be understood as follows.
When $\chi>0$, the maximum of $W_{\alpha,N}$ is $\chi$ and the minimum is $6\chi/(6+\chi)$. 
Therefore, for $\chi \in (0,(\frac{\pi}{2})^2)$, where $(\frac{\pi}{2})^2\approx 2.47$, the KA-II condition of Theorem \ref{KA} is satisfied. Moreover, when  $6\chi/(6+\chi)> (\frac{\pi}{2})^2$, that is, for $\chi>\frac{6\pi^2}{24-\pi^2}\approx 4.19$, Theorem \ref{Hannulus} guarantees the existence of an instability.
When $\chi\in (-6,0)$, the KA-I condition in Theorem \ref{KA} is satisfied. For $\chi \leq -6$, neither theorem can be applied, since the numerator of (\ref{narrowW}) vanishes at $y=\pm \sqrt{1+6/\chi}$ and $W_{\alpha,N}$ is not continuous. By numerically solving the Rayleigh equation,
\begin{eqnarray}\label{GeqNG}
\frac{d^2G}{dy^2}-k^2G-\frac{\frac{d^2U}{dy^2}}{U-c} G=0, \qquad G(-1)=G(1)=0,
\end{eqnarray}
we find that the true neutral point occurs at $\chi\approx 3.810$. Note that in the narrow-gap limit, Squire’s theorem implies that the stability is independent of the wavenumber ratio. Hence, there is no need to consider $N$ at all.

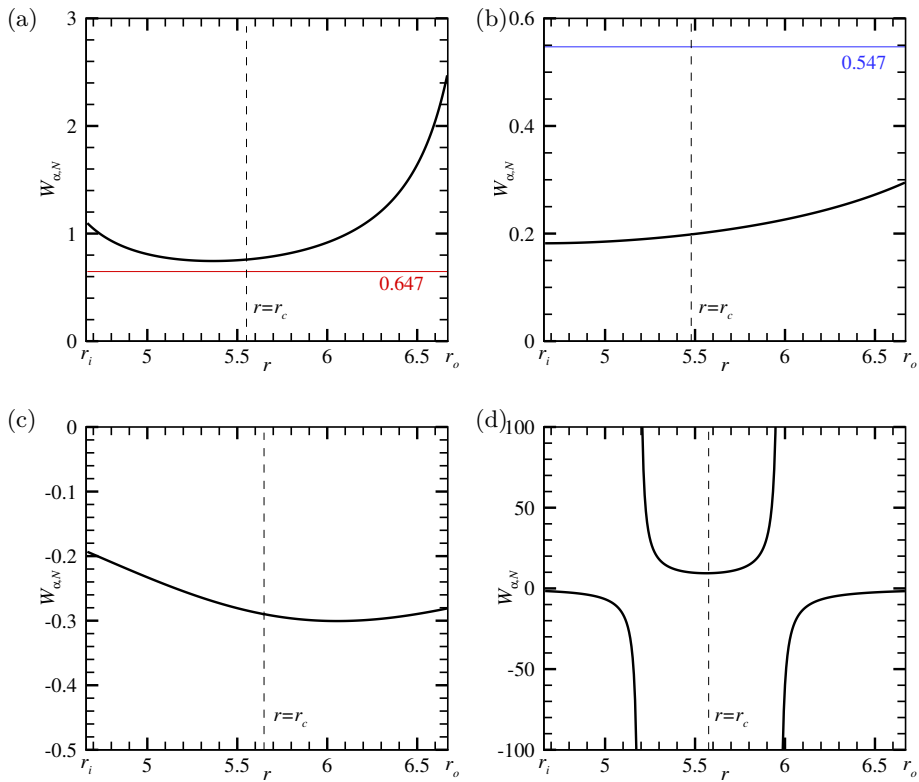
\begin{figure}
\centering
\begin{overpic}[width=0.9 \textwidth]{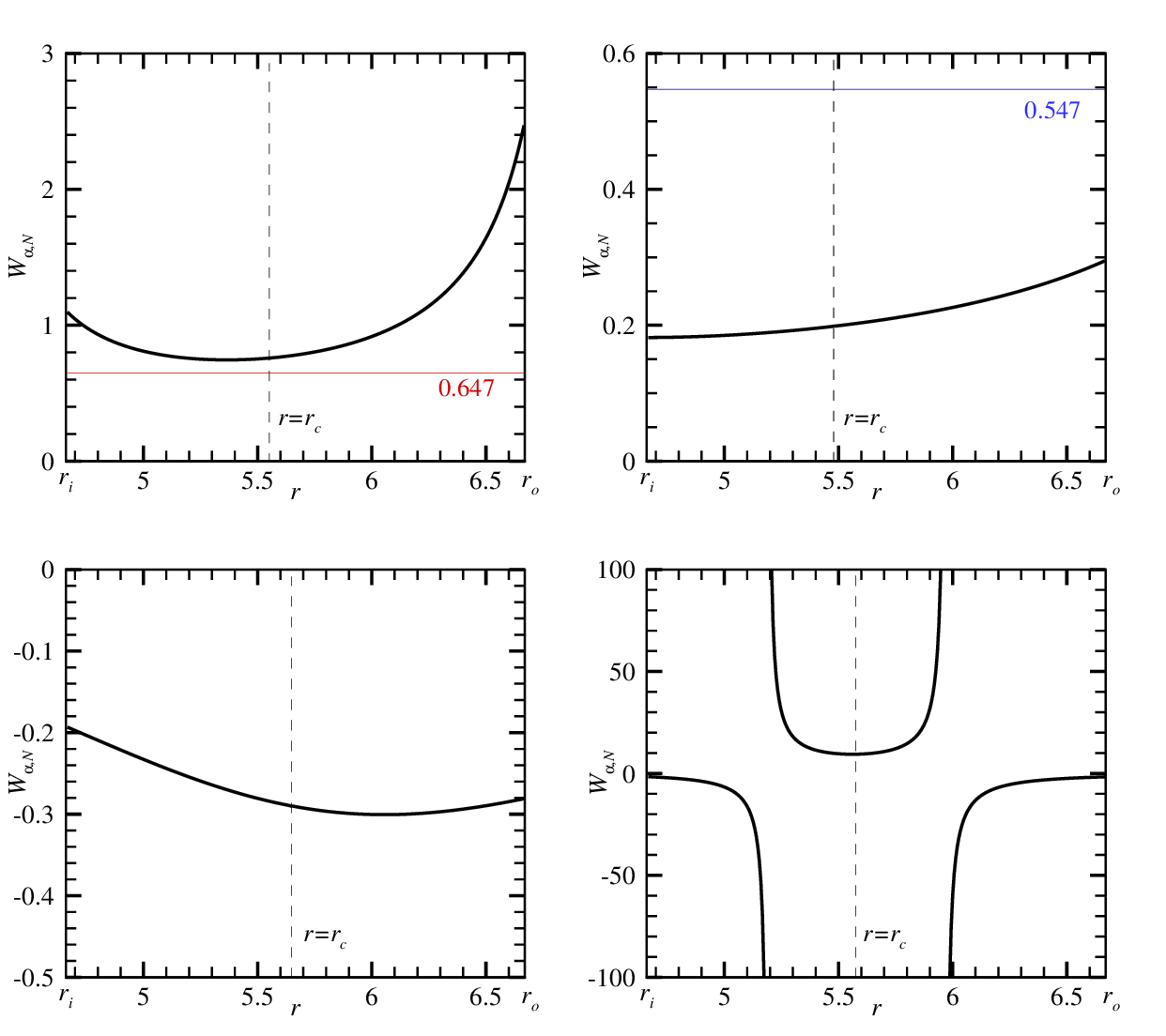}
\put(-3,83.5){(a)}
\put(48,83.5){(b)}
\put(-3,40){(c)}
\put(48,40){(d)}
\end{overpic}
\caption{
Profiles of $W_{\alpha,N}$ with $N=10$ for the annular model flow at $\eta=0.7$. The constant $\alpha$ is set equal to $U(r_c)$.
(a) $\chi=7$; (b) $\chi=1$; (c) $\chi=-1$; (d) $\chi=-7$. In panel (a), the red line shows $h$ from (\ref{hthm2}). In panel (b), the blue line shows $H$ defined in (\ref{HNHN}).} 
\label{fig:fig4}
\end{figure}
Even for $\eta<1$, the behaviour of $W_{\alpha,N}$ remains qualitatively similar to the narrow-gap case, as illustrated in figure~\ref{fig:fig4} for $\eta=0.7$.  However, the results now depend on $N$, since Squire's theorem no longer applies. In the figure, we choose $N=10$. Panel (a) illustrates the case $\chi=7$, where $W_{\alpha,N}$ surpasses the hurdle $h$ (defined in (\ref{hthm2}) and shown by the red line). 
The location $r=r_c$ indicates where
$Q'$ in (\ref{Q1annular}) vanishes.
When $\chi$ is reduced to 1, as shown in panel (b), the KA-II condition of Theorem \ref{KA} is satisfied; the blue line in the figure indicates $H$ in (\ref{HNHN}).
Panel (c) is the case where $\chi=-1$, for which $W_{\alpha,N}$ is everywhere negative, and hence the flow is stable for $N=10$ perturbations by KA-I (i.e. the result by \cite{Batchelor_Gill_1962}). Panel (d) corresponds to $\chi=-7$, which lies below the lower boundary of the stable region (blue) shown in figure~\ref{fig:fig2}. In this case, $W_{\alpha,N}$ cannot be made continuous for any choice of $\alpha$, and neither Theorem \ref{KA} nor Theorem \ref{Hannulus} can be used to determine stability. 

Figure~\ref{fig:fig4} shows only the case for $N=10$, but physically, perturbations with all values of $N$ can occur.
Therefore, to use Theorem \ref{KA} to establish stability, it is not sufficient to examine only a handful of $N$ values. 
In figure~\ref{fig:fig2}, we verified the reliability of the blue stable region using a wide range of $N$ values. 
Note that the limit $N\rightarrow \infty$ is straightforward to analyse. This means that for sufficiently large $N$, asymptotic behaviour emerges, so in practice it is sufficient to examine the conditions of the theorem up to moderately large $N$.
Likewise, at each point in the parameter plane, the value of $N$ is optimised  to maximise the red unstable region.
However, fixing $N=10$ yields almost the same outcome as that in figure~\ref{fig:fig2}. This suggests that identifying the unstable region using Theorem \ref{Hannulus} is easier than finding the stable region using Theorem \ref{KA}.

The neutral curve in figure~\ref{fig:fig2} is also obtained by optimising over the wavenumbers $k$ and $n$: above the black line an unstable mode is found for some $(k,n)$, while below it no unstable mode exists for any choice of $(k,n)$.

\begin{figure}
\centering
\begin{overpic}[width=0.9 \textwidth]{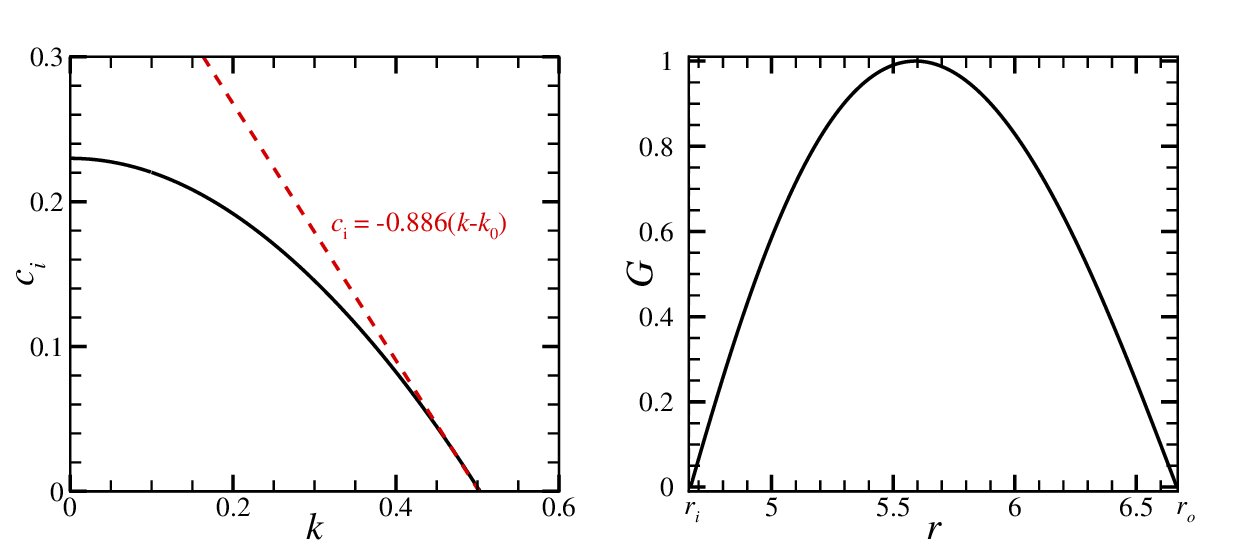}
\put(-3,40){(a)}
\put(48,40){(b)}
\end{overpic}
\caption{
Inviscid stability result for the annular model flow at $(\eta,\chi)=(0.7,7)$. (a) Imaginary part of the phase speed $c_i$ for $N=10$. The neutral point is at $k=k_0=0.502$.
The dashed red line indicates the result using (\ref{dcidk}). 
(b) Eigenfunction of the neutral mode found at $N=1/k$, $k=0.9475$ (i.e. $n=1$).
}
\label{fig:fig5}
\end{figure}

Figure~\ref{fig:fig5} shows $c_i$ obtained from numerical computations of (\ref{Geq}) for $\eta=0.7$ and $\chi=7$. Panel (a) is computed with $N=10$; hence at $k=0.1$, the mode with $n=Nk=1$ exhibits instability, as expected by Theorem~\ref{Hannulus}. 
The neutral solution at $k=0.502$ is nonphysical because $n=5.02$ is not an integer. Physically admissible neutral solutions can be obtained by reducing $N$.
For example, choosing $N=1/0.9475$ yields a neutral point at $k=k_0=0.9475$, i.e. $n=1$. Panel (b) shows the corresponding eigenfunction.
This function is free of singularities due to the regularity of $W_{\alpha,N}$. 
The red dashed line in panel (a) shows an analytic linear approximation of how $c_i$ behaves near the neutral point. 
As will be shown in section~\ref{sec:5.2} (see (\ref{eq:dcidk})), we can derive the identity
\begin{eqnarray}\label{dcidk}
\left. \frac{dc_i}{dk}\right |_{k=k_0}
=-\frac{2k_0K_i}{K_r^2+K_i^2}\int_{\Omega}rG_0^2dr,
\end{eqnarray}
where
\begin{eqnarray*}
K_r=\lim_{\epsilon\rightarrow 0}\{\int_{r_i}^{r_c-\epsilon}\frac{r^2Q'G_0^2}{(U-\alpha)^2}dr+\int^{r_o}_{r_c+\epsilon}\frac{r^2Q'G_0^2}{(U-\alpha)^2}dr\},~~
K_i=\pi \left. \left (\frac{rW_{\alpha,N}G_0^2}{|U'|} \right) \right|_{r=r_c}
\end{eqnarray*}
can be found using the neutral wavenumber $k_0$, the regular neutral eigenfunction $G_0$, and $\alpha=U(r_c)$.
The right hand side of (\ref{dcidk}) is evaluated as $-0.886$, which gives the slope of the red dashed line. As similar results hold for $N\approx 1/0.9475$, one can conclude that a slight increase of $N$ from its neutral value necessarily results in the emergence of an unstable mode with $n=1$.
The observation here highlights the strategy to be used in section~\ref{sec:5.2} for proving Theorem~\ref{Hannulus}.

Along the neutral curve in figure~\ref{fig:fig2}, the eigenfunction is regular.
However, when $W_{\alpha,N}$ is singular (as in figure~\ref{fig:fig4}-(d)), neutral solutions with singularities at the critical levels (i.e. locations where $U=c$) may arise. 
In the parameter plane shown in figure~\ref{fig:fig2}, such singular modes indeed occur when $\chi$ lies slightly below the lower boundary of the stable region. A detailed check in the narrow-gap limit confirmed that the neutral point lies very close to $\chi=-6$ (see figure~\ref{fig:fig3}). The computation was performed using 30000 grid points in order to capture the nearly singular eigenfunctions with small $c_i$. 
Likewise, the blue lower stability boundary in figure \ref{fig:fig2} may lie very close to the neutral curve, but we do not go into further detail. 
Note that such nearly singular unstable modes cannot be captured by Theorem~\ref{Hannulus} and are therefore outside the scope of the present study.

\begin{figure}
\centering
\begin{overpic}[width=0.9 \textwidth]{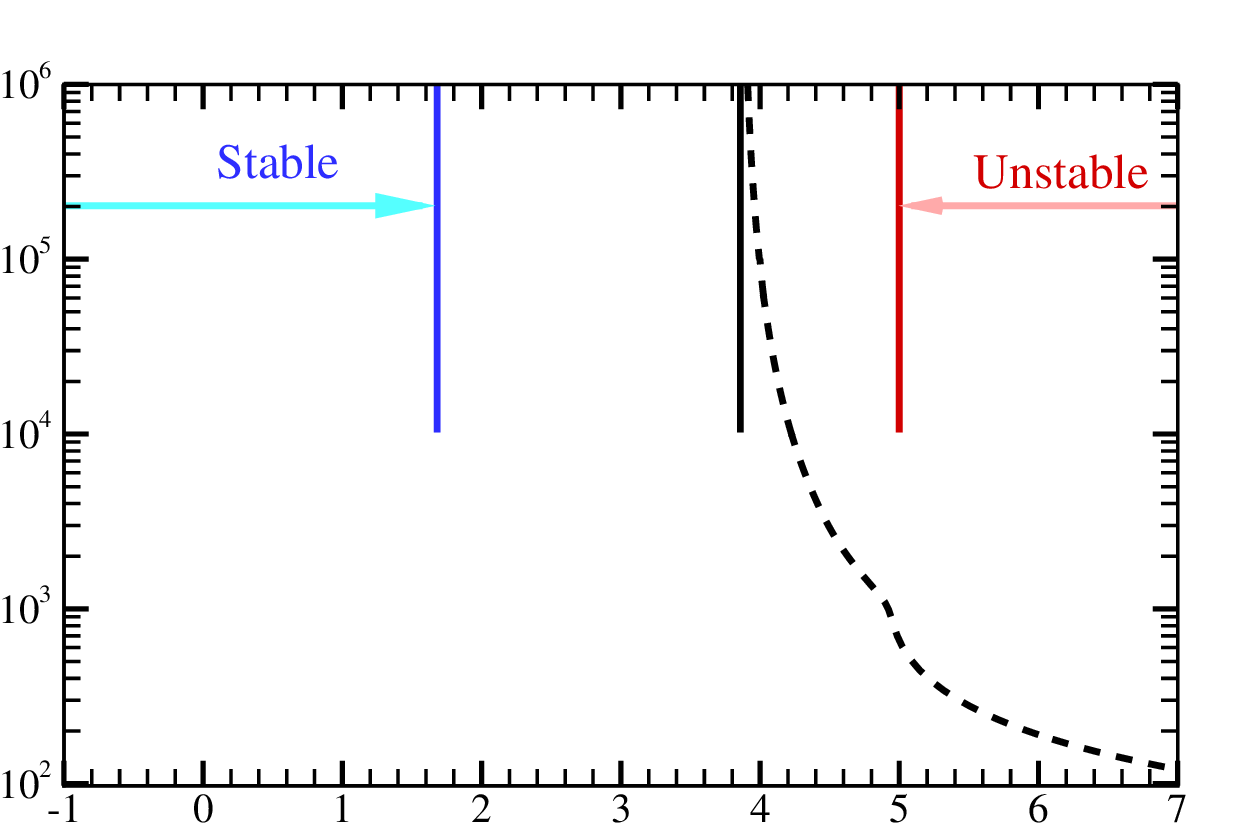}
\put(-3,30){\begin{turn}{90}{$Re$}\end{turn}}
\put(49,-2){{$\chi$}}
\put(52,29){{$\chi$=3.86}}
\put(47,25){{Inviscid limit}}
\put(30,29){{$\chi$=1.68}}
\put(70,29){{$\chi$=5.00}}
\end{overpic}
\caption{
Comparison between the viscous and inviscid stability analyses for the annular model flow at $\eta=0.7$. The dashed line represents the neutral curve obtained from the viscous stability problem (\ref{linNS}), covering all physically possible wavenumbers. The blue, black, and red vertical lines correspond to the inviscid stability results shown in figure~\ref{fig:fig2}.
}
\label{fig:fig6}
\end{figure}
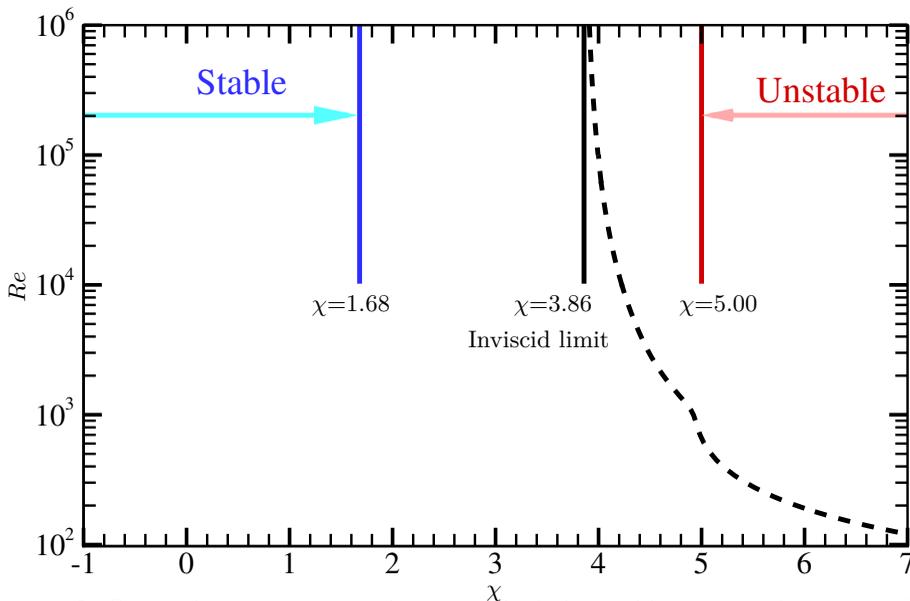
Figure~\ref{fig:fig6} confirms the above inviscid stability results through computations of the linearised Navier-Stokes equations (\ref{linNS}). The dashed line indicates the neutral curve obtained by optimising over wavenumbers $k$,$n$. The most unstable mode is always $n=1$, which asymptotes to the inviscid result $\chi=3.86$ as $Re$ increases. The stability for $Re>10^4$ is relatively well captured by Theorems~\ref{KA} and \ref{Hannulus}.

\subsection{Flow through a pipe}\label{sec:4.2}

\begin{figure}
\centering
\begin{overpic}[width=0.45 \textwidth]{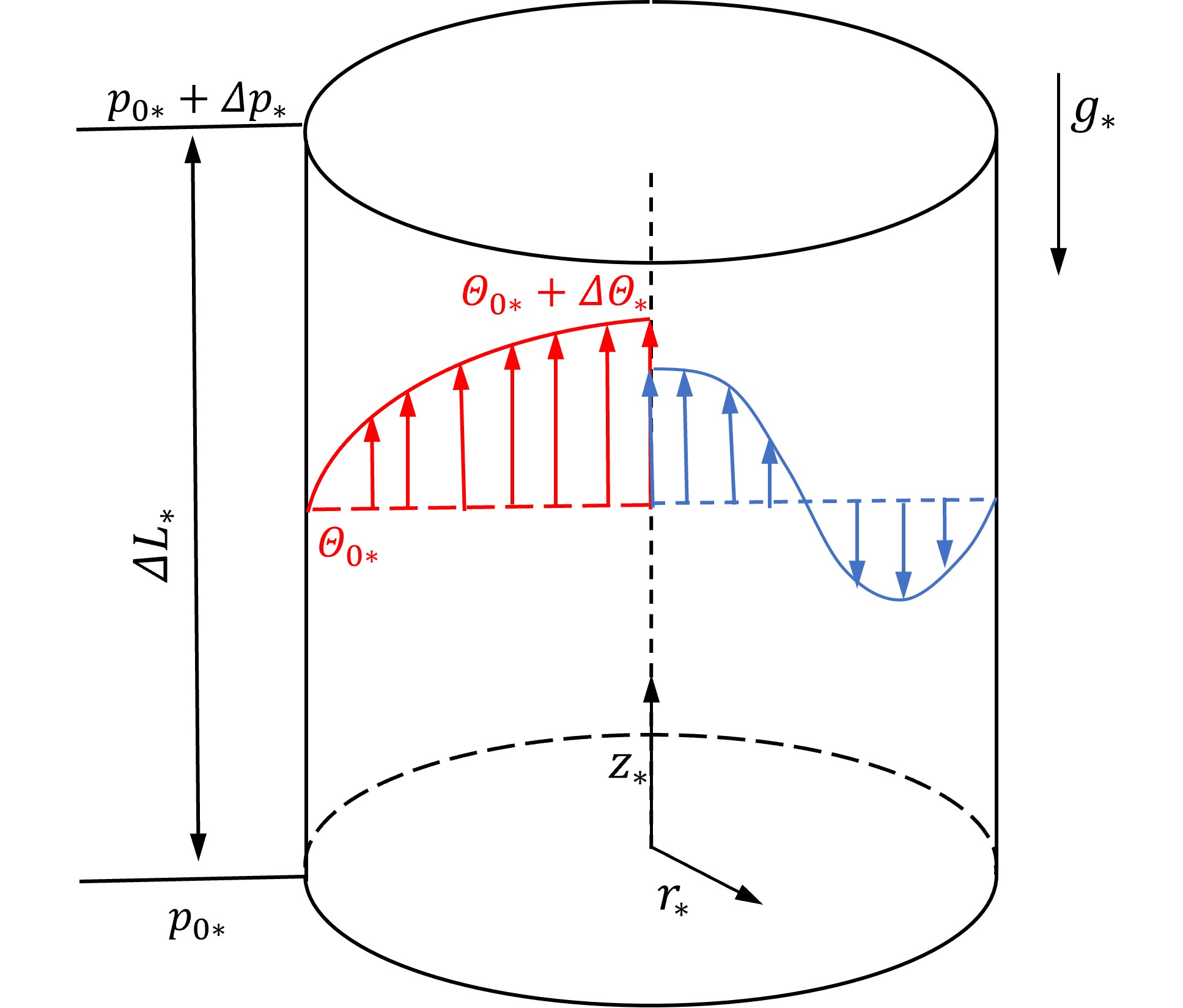}
\put(-2,80){(a)}
\end{overpic}
\begin{overpic}[width=0.45 \textwidth]{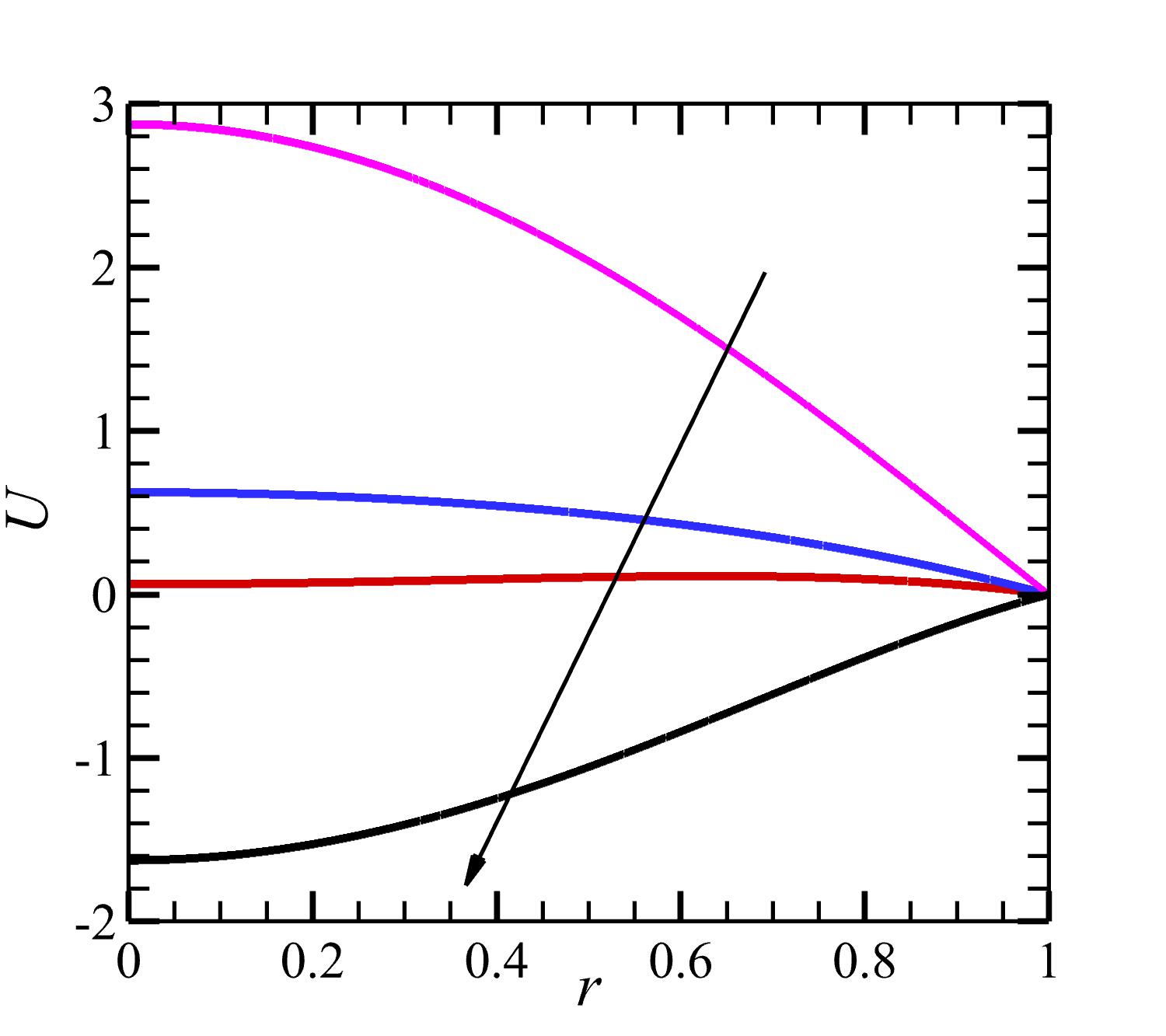}
\put(-3,80){(b)}
\put(50,70){Increasing $\chi$}
\end{overpic}
\caption{
The model flow used in section~\ref{sec:4.2}. 
(a) Schematic of the model flow in the dimensional cylindrical coordinates $(r_*,\varphi,z_*)$.
(b) The base flow $U(r)$ given in (\ref{PipeU}) for $\chi=-10,5,7,14$. 
}
\label{fig:fig7}
\end{figure}

In this section, we consider vertically oriented Hagen–Poiseuille flow with homogeneous internal heating (figure~\ref{fig:fig7}-(a), see \cite{Senoo2012}, \cite{Marensi_He_Willis_2021} also). We choose the pipe radius $d_*$ as the length scale, and the centreline velocity of the laminar Hagen–Poiseuille flow $u_*$ as the velocity scale.
For the temperature scale, we use the difference $\Delta\theta_*$ between the base temperature at the centreline, $\theta_{0*}+\Delta\theta_*$, and that at the wall, $\theta_{0*}$.
The velocity and temperature scales are related to the dimensional axial pressure gradient $\Delta p_*/L_*$ and the dimensional internal heat source $q_*$ as follows.
\begin{eqnarray}
u_*=\frac{d_*^2}{4\mu_*}\frac{\Delta p_*}{L_*},\qquad \theta_*=\frac{d_*^2q_*}{4\kappa_*}.
\end{eqnarray}
Here, $\mu_*$ is the dynamic viscosity and $\kappa_*$ is the thermal diffusivity of the fluid.

From the Boussinesq-approximated Navier–Stokes equations, the equations for the axial base velocity $U(r)$ and the base temperature $\Theta(r)$ are obtained as
\begin{eqnarray}
\Theta''+r^{-1}\Theta'=-4,\\
U''+r^{-1}U'=\chi \Theta-4.
\end{eqnarray}
The solutions of the above equations, satisfying the boundary conditions 
\begin{eqnarray*}
U=0, ~~\Theta=0\qquad \text{at}\qquad r=1.
\end{eqnarray*}
and the centreline regularity, are 
\begin{eqnarray}
\Theta=(1-r^2),\\
U=(1-r^2)-\frac{\chi }{16}(1-r^2)(3-r^2).\label{PipeU}
\end{eqnarray}
The base flow profiles for selected values of $\chi$ are shown in figure~\ref{fig:fig7}-(b). 
An inflection point appears when $\chi$ is in a certain range, and by analogy with Rayleigh's theorem, \cite{Senoo2012} speculated that this could be a possible cause of instability. However, the mathematics of inviscid instability is not that simple.

\begin{figure}
\centering
\begin{overpic}[width=0.8 \textwidth]{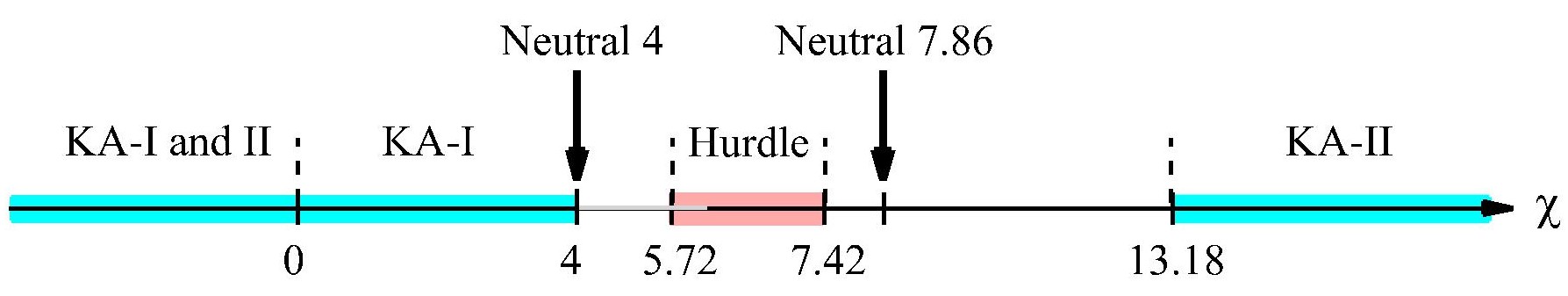}
\end{overpic}
\caption{
Stability diagram of the pipe model flow (figure~\ref{fig:fig7}-(a)).
The eigenvalue problem (\ref{Geq}) indicates the presence of unstable modes for $\chi \in (4,7.86)$.
The grey line shows that the profile $W_{\alpha,N}$ becomes singular when $\chi \in [4,6]$. 
}
\label{fig:fig8}
\end{figure}

Substituting the base flow $U$ into the definition of $Q$ (see (\ref{Geq})) and differentiating, we obtain
\begin{eqnarray}\label{Q1pipe}
Q'=r\frac{N^2(4-\chi(1-r^2))+\frac{\chi}{2}r^4}{(N^2+r^2)^2}.
\end{eqnarray}
It is easy to see that $Q'$ vanishes at $r=r_c$, where
\begin{eqnarray}
r_c=N\sqrt{\sqrt{1+\frac{2}{N^2}(1-\frac{4}{\chi})}-1 }. \label{piperc}
\end{eqnarray}
Clearly, the axisymmetric mode is stable, since for $N=0$, the function $Q'$ does not change sign in the domain (the KA-I condition is satisfied, see the remark in section~\ref{sec:3}). 
Even for non-axisymmetric modes, the parameter range $\chi \in (0,4)$ remains stable, as no real $r_c$ exists. Determining inviscid stability for other values of $\chi$ is not trivial, but by applying Theorems~\ref{KA} and \ref{Hannulus}, we can find the results shown in figure~\ref{fig:fig8}. 

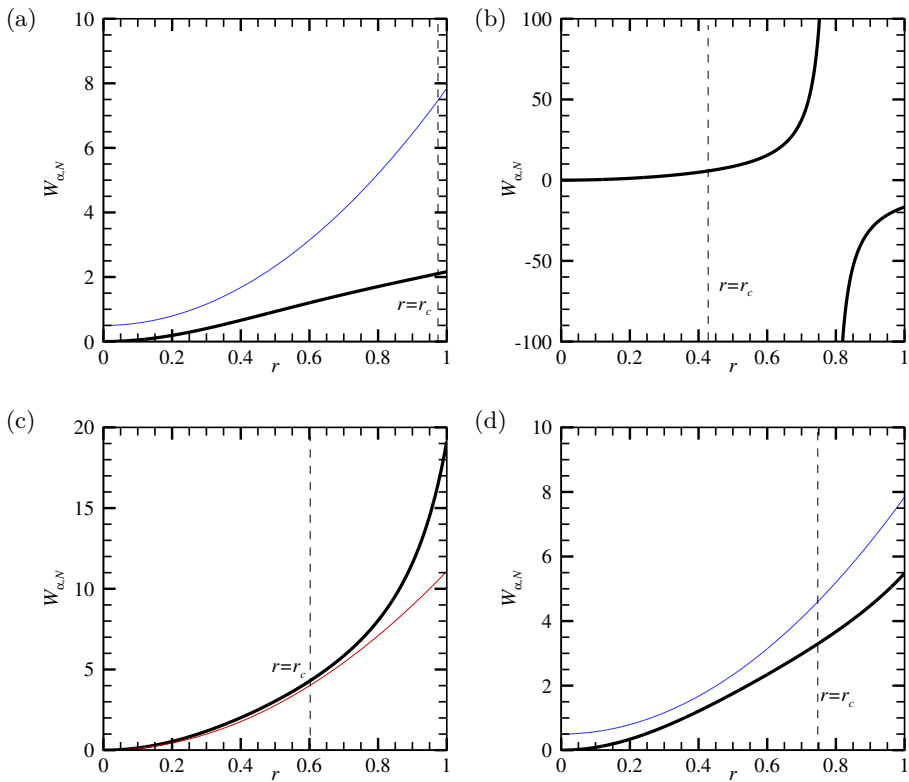
\begin{figure}
\centering
\begin{overpic}[width=0.9 \textwidth]{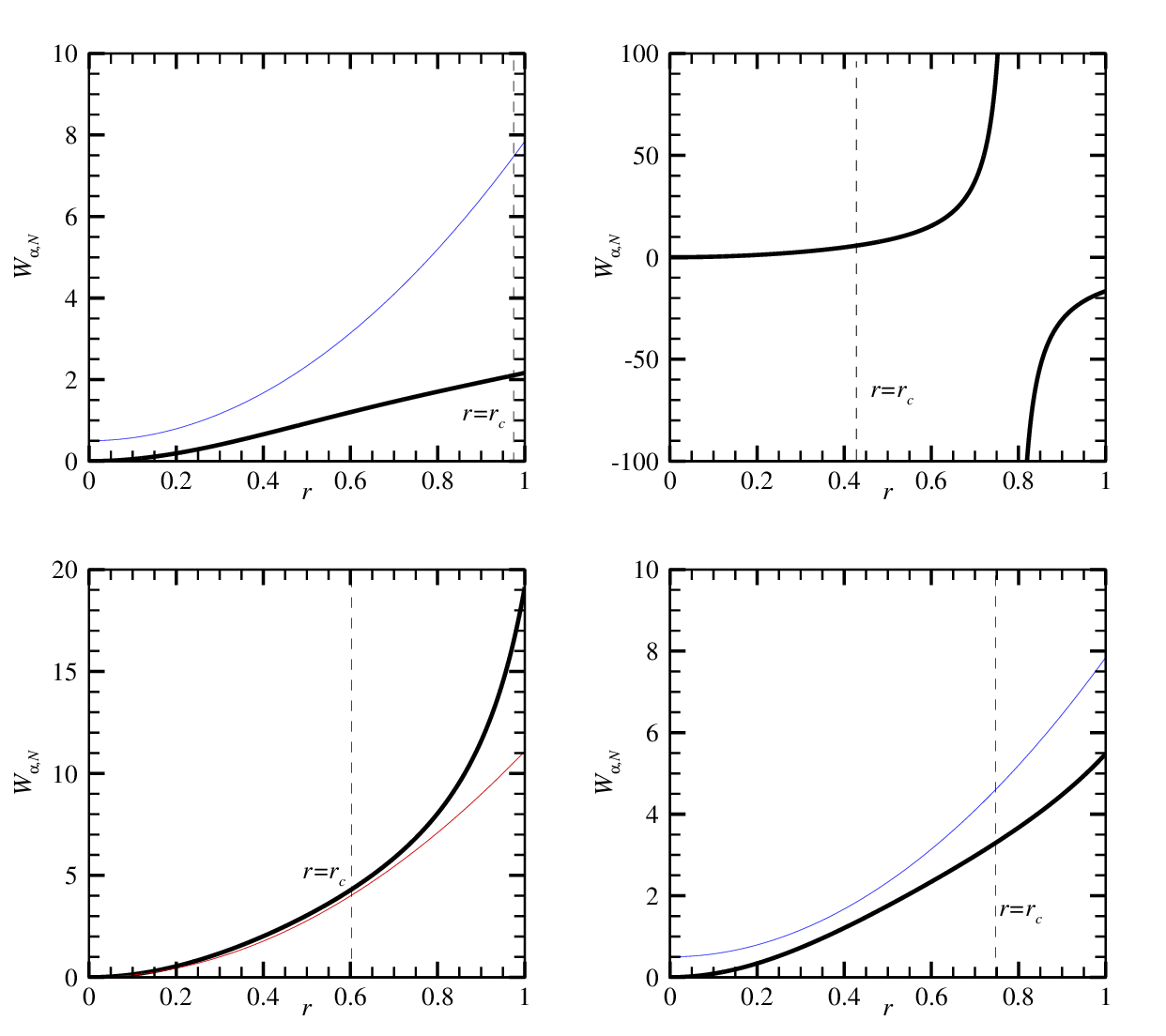}
\put(-3,83.5){(a)}
\put(48,83.5){(b)}
\put(-3,40){(c)}
\put(48,40){(d)}
\end{overpic}
\caption{
Profiles of $W_{\alpha,N}$ with $\alpha=U(r_c)$ and $N=1$ for the pipe model flow. (a) $\chi=-10$; (b) $\chi=5$; (c) $\chi=7$; (d) $\chi=14$. 
In panel (c), the red line shows $h$ from (\ref{hthm3}). In panels (a) and (d), the blue line shows $H$ defined in (\ref{HNHN2}). 
}
\label{fig:fig9}
\end{figure}
When $\chi$ is negative, either KA-I or KA-II holds.
Figure~\ref{fig:fig9}-(a) shows the profile of $W_{\alpha,N}$ for $\chi=-10$ and $N=1$. This is an example in which the KA-II condition defined in (\ref{HNHN2})) is satisfied;  
$W_{\alpha,N}$ is positive everywhere and lies below $H$ indicated by the blue line. 

\begin{figure}
\centering
\begin{overpic}[width=0.9 \textwidth]{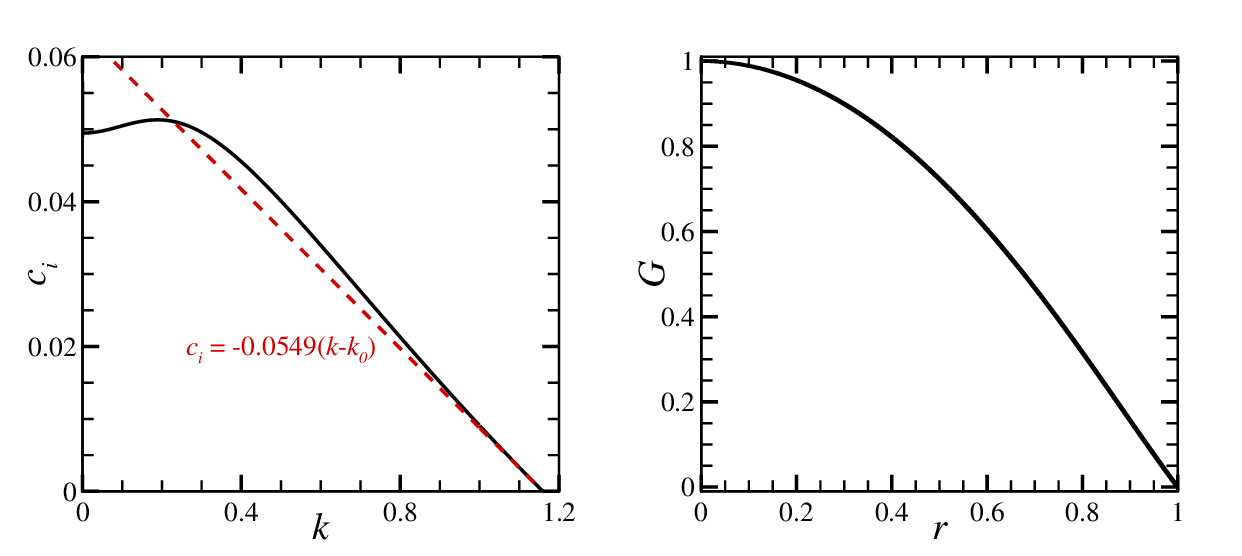}
\put(-3,40){(a)}
\put(48,40){(b)}
\end{overpic}
\caption{
Inviscid stability result for the pipe model flow at $\chi=7$. (a) Imaginary part of the phase speed $c_i$ for $N=1$. The neutral point is at $k=k_0=1.159$. The dashed red line indicates the result using (\ref{dcidk}).
(b) Eigenfunction of the neutral mode found at $N=1/k$, $k=1.46$ (i.e. $n=1$).
}
\label{fig:fig10}
\end{figure}
For $\chi \in [5.72,7.42]$, there exist $N>0$ and $p\geq 2$ such that $W_{\alpha,N}$ exceeds the hurdle (\ref{hthm3}).  
Figure~\ref{fig:fig9}-(c) shows the profile of $W_{\alpha,N}$ for $\chi=7$ and $N=1$. In this figure, the hurdle for $p=2$ is indicated by the red line. Thus, by Theorem~\ref{Hannulus}, the flow is unstable. Note that a horizontal hurdle, as used in the annular case, does not work well for the pipe flow since near $r=0$, $W_{\alpha,N}$ behaves like $r^2$. 
Figure~\ref{fig:fig10}-(a) shows the eigenvalue analysis of (\ref{Geq}) for $\chi=7$ and $N=1$. This calculation uses method (B) of section~\ref{sec:2.2}, with the condition $G'(\epsilon)=0$. At $k=1$, we indeed have an unstable mode with $n=1$. 

The red dashed line in figure~\ref{fig:fig10}-(a) corresponds 
to results similar to those seen in figure~\ref{fig:fig5}-(a). The slope $-0.0549$ is obtained by the neutral mode using (\ref{dcidk}). By decreasing the value of $N$ from 1 to $1/k$ with $k=1.46$, a neutral mode with $n=1$ is obtained. This neutral mode is regular as shown in figure \ref{fig:fig10}-(b). We confirmed that the methods (A) and (B) introduced in section~\ref{sec:2.2} produce  eigenfunctions that are graphically indistinguishable.

By computing neutral points using (\ref{Geq}) for all physically admissible wavenumbers, it is found that a neutral point occurs at $\chi=7.86$ (see figure~\ref{fig:fig8}). This value is reasonably close to the unstable region predicted by Theorem~\ref{Hannulus}. 
The eigenvalue problem (\ref{Geq}) does not produce unstable modes for $\chi>7.86$. 
This stabilisation can be detected by KA-II when $\chi>13.18$.
For example, figure~\ref{fig:fig9}-(d) shows the case $\chi=14$, $N=1$, where $W_{\alpha,N}$ lies below the blue line $H$ determined by (\ref{HNHN2}). 

Another neutral point is expected to exist between the blue stability region  and the red instability region in figure~\ref{fig:fig8}. The computations of (\ref{Geq}) indicates that instability sets in when $\chi$ slightly exceeds 4, implying that the stability boundary $\chi=4$ by Theorem~\ref{KA} sharply predicts the neutral point. The eigenfunction corresponding to this neutral point is singular. Hence, it makes sense that the boundary of the unstable region by Theorem~\ref{Hannulus} is not sharp. Careful observation of $W_{\alpha,N}$ shows that it becomes singular for some $N$ when $\chi \in [4,6]$.
An example of such a singular case is shown in figure~\ref{fig:fig9}-(b), obtained for $\chi=5$ and $N=1$.

\begin{figure}
\centering
\begin{overpic}[width=0.9 \textwidth]{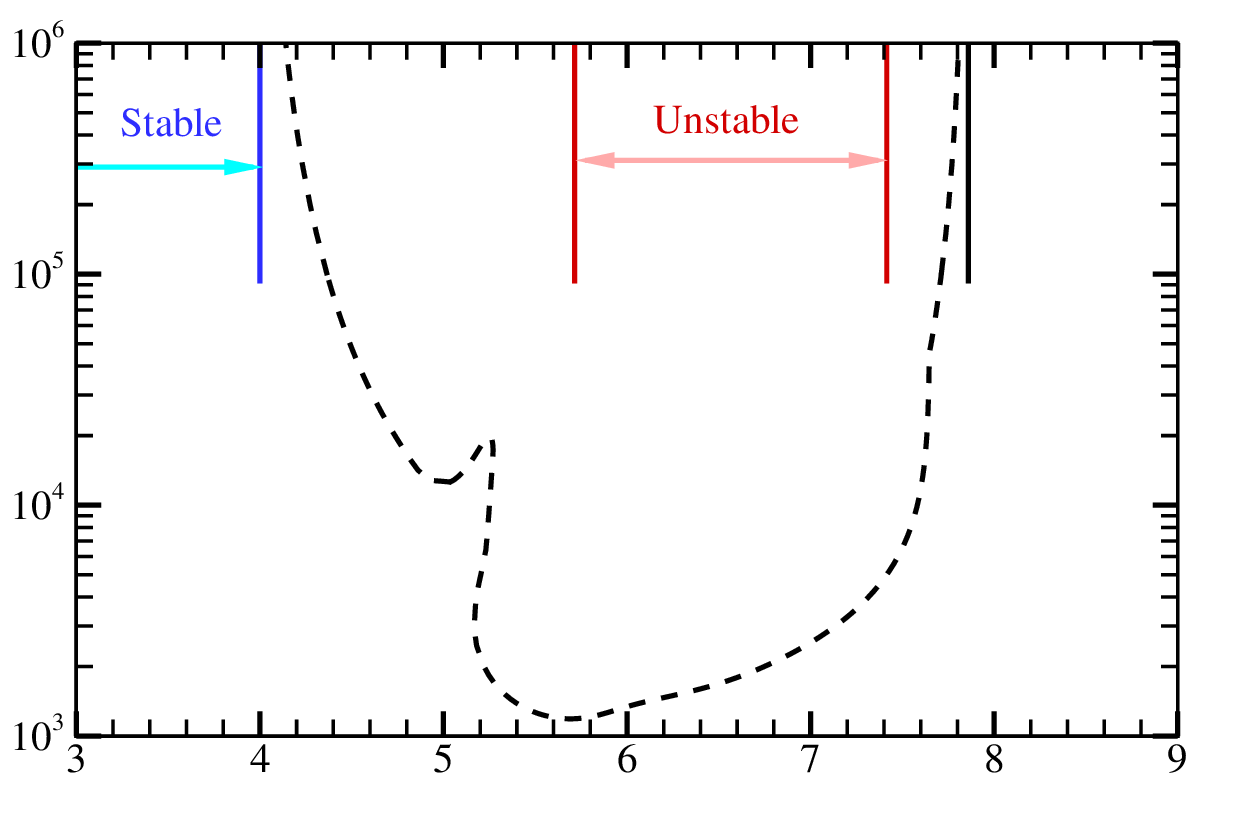}
\put(0,33){\begin{turn}{90}{$Re$}\end{turn}}
\put(50,2){{$\chi$}}
\put(77,41){{$\chi$=7.86}}
\put(77,36){{Inviscid limit}}
\put(64,41){{$\chi$=7.42}}
\put(41,41){{$\chi$=5.72}}
\put(18,41){{$\chi$=4}}
\put(12,36){{Inviscid limit}}
\end{overpic}
\caption{
Comparison between the viscous and inviscid stability analyses for the pipe model flow. 
The dashed line represents the neutral curve obtained by (\ref{linNS}) varying wavenumbers. 
The blue, black, and red vertical lines correspond to the inviscid stability results shown in figure~\ref{fig:fig8}.
}
\label{fig:fig11}
\end{figure}
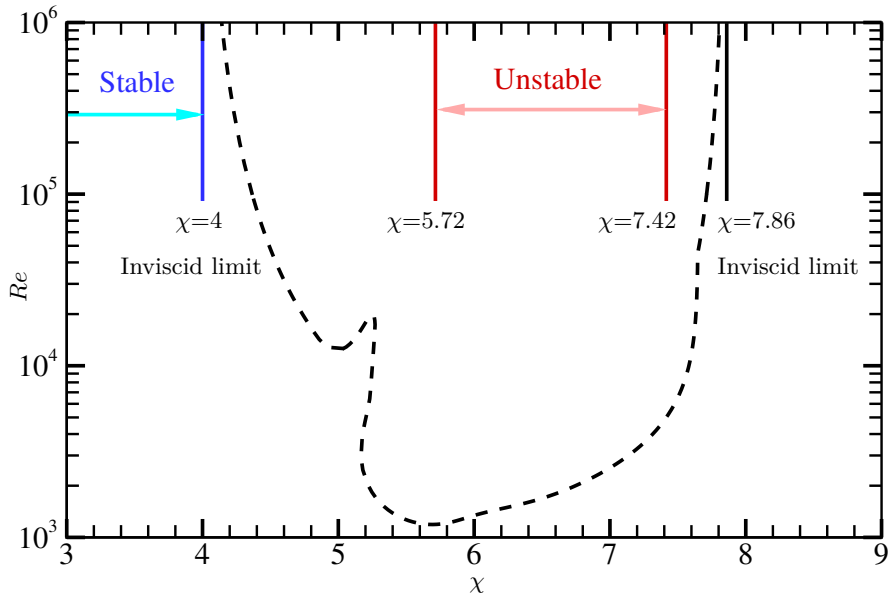
Figure~\ref{fig:fig11} verifies the above inviscid results using viscous computations.
The dashed line  represents the envelope of the neutral curves obtained from the linearised Navier-Stokes equations (\ref{linNS}) over a range of wavenumbers.
This curve is determined by the $n=1$ mode and
asymptotes to the inviscid neutral points at $\chi=4$ and $7.86$ for large Reynolds numbers. 
Recall that for $\chi \in [4,6]$, instability may arise from the singular neutral mode, whereas for $\chi \in [5.72,7.42]$, Theorem~\ref{Hannulus} guarantees the existence of regular neutral modes.
The neutral curve found by the viscous analysis features two humps, representing two unstable modes, which correspond to a singular mode and a regular mode in the inviscid analysis.

\section{Mathematical proofs} \label{sec:5}

Here, we prove Theorems \ref{KA} and ~\ref{Hannulus} stated in section~\ref{sec:3}.
For definiteness, we introduce the function space $H_0^1$, following the standard notation in functional analysis. For $f \in H_0^1$, both $\int_{\Omega} (f')^2 r,dr$ and $\int_{\Omega} f^2 r,dr$ are finite, and the boundary conditions are satisfied. Specifically, in the annular case, $f(r_i)=f(r_o)=0$, while in the pipe case, $f(1)=0$.
Recall that the pipe solution also satisfy the centreline regularity condition (\ref{taylor}). If $G$ satisfies (\ref{Geq}) in the classical sense, then $G \in H_0^1$.

\subsection{Proof of Theorem 1}\label{sec:5.1}

Following \cite{Batchelor_Gill_1962}, we multiply (\ref{Geq}) by the complex conjugate of $rG$ and integrate over the entire domain $\Omega$.
The real and imaginary parts of the integral are obtained as
\begin{eqnarray}
\int_{\Omega} \{\frac{r}{N^2+ r^2}|(rG)'|^2 +\frac{k^2}{r}|rG|^2-\frac{Q'(U-c_r)}{|U-c|^2}|rG|^2 \}dr=0,\label{realint}\\
c_i\int_{\Omega} \frac{Q'}{|U-c|^2}|rG|^2dr=0,\label{imagint}
\end{eqnarray}
respectively. The boundary terms arising from integration by parts vanish in all cases we consider.

Now, we assume $c_i\neq 0$ 
and show that the KA-I assumption leads to a contradiction. This can be found in the same manner as Fjortoft’s analysis for parallel flows.
By taking a suitable linear combination of the two integrals (\ref{realint}) and (\ref{imagint}), we obtain
\begin{eqnarray}\label{intka1}
\int_{\Omega} \{\frac{r}{N^2+r^2}|(rG)'|^2 +\frac{k^2}{r}|rG|^2-W_{\alpha,N}\frac{(U-\alpha)^2}{|U-c|^2}\frac{|rG|^2}{r} \}dr=0,
\end{eqnarray}
where $\alpha \in \RealN$ is arbitrary.
If $W_{\alpha,N}=rQ'/(U-\alpha)\leq 0$ for all $r \in \Omega$, the above integral cannot be satisfied for non-trivial $G$.
The analysis here 
is identical to that in \cite{Batchelor_Gill_1962} but included for completeness.

KA-II stability is usually derived by defining a Hamiltonian. However, in our case, an equivalent condition can be obtained in a more straightforward manner. 
We first set $\alpha=2c_r-\beta$ to rewrite (\ref{intka1}) in the form
\begin{eqnarray}\label{Zeq}
\int_{\Omega} \{\frac{r}{N^2+ r^2}|(rG)'|^2 +\frac{k^2}{r}|rG|^2-W_{\beta,N}Z\frac{|rG|^2}{r} \}dr=0,
\end{eqnarray}
where $Z=\frac{(U-c_r)^2-(c_r-\beta)^2}{(U-c_r)^2+c_i^2}< 1$. The choice of $\beta \in \RealN$ is arbitrary; by assumption, it can be adjusted so that $W_{\beta,N}\in [0,H(r))$ 
for $r\in \Omega$. From this point onward, the analysis must be treated separately for annular and pipe flows.

For the annular problem, there is a positive constant $\kappa_N^2$ depending on $N$ such that
\begin{eqnarray}
\int_{r_i}^{r_o} \frac{r}{N^2+ r^2}|(rf)'|^2 dr\geq \kappa_N^2 \int_{r_i}^{r_o} r^{-1}|rf|^2dr,\label{fineq}
\end{eqnarray}
for all $f(r) \in H_0^1$; this is a generalisation of Poincar\'e's inequality. 
Using the non-negativity of $W_{\beta,N}$ and (\ref{fineq}) in (\ref{Zeq}), we obtain
\begin{eqnarray}\label{Zeq2}
0> \int_{r_i}^{r_o} \left (\kappa_N^2 +k^2-W_{\beta,N} \right ) \frac{|rG|^2}{r} dr.
\end{eqnarray}
Computation in Appendix~\ref{sec:AppA1} shows that 
\begin{eqnarray}\label{kappaN}
\kappa_N^2=\frac{(\pi/\ln \eta)^2}{N^2+r_o^2}, 
\end{eqnarray}
and with this expression, (\ref{HNHN}) becomes
\begin{eqnarray}
H=\left \{
\begin{array}{c}
\kappa_N^2+\frac{1}{N^2}\qquad \text{if} \qquad N\geq 1,\\
\kappa_0^2\qquad \text{if} \qquad N= 0.
\end{array}
\right .
\end{eqnarray}
If $n=0$ (i.e. $N=0$) and $W_{\beta,0}<\kappa_0^2$, clearly the inequality (\ref{Zeq2}) cannot be satisfied by non-trivial $G$.
When $n\neq 0$ (i.e. $N>0$), we can use $k^2N^2=n^2\geq 1$.
Thus, in this case, when
$W_{\beta,N}<\kappa_N^2+\frac{1}{N^2}$ the inequality (\ref{Zeq2}) cannot be satisfied.

For the pipe problem, different inequalities must be applied separately to the axisymmetric case ($n=0$) and the non-axisymmetric cases ($n\geq 1$).
If $n=0$, we use the fact that
\begin{eqnarray}
\int_{0}^{1} \frac{1}{r}|(rf)'|^2 dr\geq j_{1,1}^2 \int_{0}^{1} \frac{1}{r}|rf|^2dr,\label{ineq1}
\end{eqnarray}
for all $f\in H_0^1$ satisfying the regularity condition (see Appendix~\ref{sec:AppA2}).
Then (\ref{ineq1}) and the integral (\ref{Zeq}) imply that
\begin{eqnarray}
0> \int_0^1 \left (j_{1,1}^2 +k^2-W_{\beta,0} \right ) \frac{|rG|^2}{r} dr.
\end{eqnarray}
Clearly this does not happen when $W_{\beta,0}\leq j_{1,1}^2$.

If $n\geq 1$, 
we can use the following inequality, which holds for all regular $f\in H_0^1$ (see Appendix~\ref{sec:AppA3}):
\begin{eqnarray}\label{ineq2}
\int^1_0 \{r|(rf)'|^2 +\frac{n^2}{r}|rf|^2\}dr\geq j_{1,1}^2 \int^1_0 r|rf|^2dr.
\end{eqnarray}
The integral (\ref{Zeq}) then yields the inequality
\begin{eqnarray}
0> \int^1_0 \{\frac{r}{N^2+ 1}|(rG)'|^2 +\frac{n^2}{rN^2}|rG|^2-W_{\beta,N}\frac{|rG|^2}{r} \}dr\nonumber \\
=
\int^1_0 \{\frac{r}{N^2+ 1}|(rG)'|^2 +\frac{n^2}{r(N^2+1)}|rG|^2\}dr\nonumber\\
+\int^1_0\{
\frac{n^2}{rN^2(N^2+1)}|rG|^2-W_{\beta,N}\frac{|rG|^2}{r} \}dr\nonumber \\
\geq
\int^1_0\{ \frac{j_{1,1}^2r^2}{N^2+ 1}
 +
\frac{1}{N^2(N^2+1)}-W_{\beta,N}\}\frac{|rG|^2}{r} dr,
\end{eqnarray}
which is impossible when $W_{\beta,N}\leq (j_{1,1}^2 r^2+N^{-2})/(N^2+1)$ for all $r \in (0,1)$.

\subsection{Proof of Theorem 2: step (i)} \label{sec:5.2}

To carry out step (i) described in section~\ref{sec:introduction}, we first note that a neutral solution satisfies the integral equation
\begin{eqnarray}
\frac{\int_{\Omega}\{ \frac{r}{N^2+ r^2}|(rG)'|^2 -W_{c,N}\frac{1}{r}|rG|^2 \} dr}{\int_{\Omega} \frac{1}{r}|rG|^2 dr}=-k^2\label{intneu}
\end{eqnarray}
with $c \in \RealN$.
This observation motivates us to define a functional $R_{\alpha,N}: H_0^1 \rightarrow \RealN$ depending on $\alpha, N\in \RealN$ by
\begin{eqnarray}\label{Rayq}
R_{\alpha,N}(\phi)=\frac{\int_{\Omega}\{ \frac{r}{N^2+ r^2}|(r\phi)'|^2 -W_{\alpha,N}\frac{1}{r}|r\phi|^2 \} dr}{\int_{\Omega} \frac{1}{r}|r\phi|^2 dr}.
\end{eqnarray}
Then, (\ref{intneu}) can be compactly written as $R_{c,N}(G)=-k^2$. A function that satisfies this equation with real $k$ and $c$ may correspond to a neutral solution, but at this stage, it is not obvious whether such a solution exists.

The key mathematical fact we use is that when  $W_{\alpha,N}$ is continuous, the minimum of the Rayleigh quotient (\ref{Rayq}),  
\begin{eqnarray}\label{RayQ}
\lambda_0=\min_{\phi \in H_0^1} R_{\alpha,N}(\phi), 
\end{eqnarray}
exists, thanks to the completeness of $H_0^1$. The minimiser $\phi_0$  satisfies the Euler-Lagrange equation
\begin{eqnarray}\label{SLSL}
\{\frac{r}{N^2+r^2}(r\phi)' \}'+(\lambda+W_{\alpha,N})\phi=0,
\end{eqnarray}
with $\lambda=\lambda_0$ and $\phi=\phi_0$. 
This equation has the same form as (\ref{Geq}).
Hence, by setting $G=\phi_0$, $c=\alpha$, and $k=\sqrt{-\lambda_0}$, a neutral solution is constructed, although physically $\lambda_0$ must be negative at least.
The remaining task is to determine the condition under which $\lambda_0$ becomes sufficiently small. For this purpose, we will introduce  appropriate trial functions in section~\ref{sec:5.4}.

Here we remark that for the annular domain, (\ref{SLSL}) together with the boundary conditions forms a regular Sturm–Liouville problem, which, as is well-known, 
admits countably many real eigenvalues that can be ordered as $\lambda_0 < \lambda_1 < \dots < \lambda_m < \dots$, with $\lambda_m \to \infty$ as $m \to \infty$. The minimum of the Rayleigh quotient corresponds to the smallest eigenvalue $\lambda_0$. The associated minimiser $\phi_0$ is real and has no zeros in $\Omega$. This latter property follows from Sturm's oscillation theorem and plays an important role in step (ii). 

For the pipe case, however, (\ref{SLSL}) is a singular Sturm–Liouville problem; readers interested in the details of regular and singular Sturm–Liouville problems are referred to the classical textbooks by \cite{Courant_Hilbert}, \cite{Titchmarsh_1962} and \cite{Birkhof_1989} or the more modern treatment by \cite{Zettl2005}. Singular cases can be further classified; our pipe problem belongs to the limit-point non-oscillatory case, where the eigenfunction behaves relatively well near $r=0$.

That said, there is no need to distinguish between the annular and pipe cases in the proof, except when constructing the trial function that determines $h$ in the theorem in section~\ref{sec:5.4}. Recall that in method (B) of section~\ref{sec:2.2}, we considered an annular problem with a thin virtual inner cylinder of radius $\epsilon>0$. For this augmented pipe problem, the discussions in sections~\ref{sec:5.2} and \ref{sec:5.3} remain applicable. We then expect that the neutral or unstable solutions constructed there will converge to those of the pipe problem as $\epsilon \rightarrow 0$. We have already demonstrated numerically that this expectation holds in section~\ref{sec:4.2}; for a mathematical justification, see the textbook by \cite{Titchmarsh_1962}, for example.

\subsection{Proof of Theorem 2: step (ii)} \label{sec:5.3}
For both the annular and pipe cases, the assumptions of Theorem \ref{Hannulus} are necessary to guarantee the existence of a neutral solution using specific trial functions $\psi$. 
The proof is deferred to section~\ref{sec:5.4}, while step (ii) is presented first.

\def\Xint#1{\mathchoice
   {\XXint\displaystyle\textstyle{#1}}%
   {\XXint\textstyle\scriptstyle{#1}}%
   {\XXint\scriptstyle\scriptscriptstyle{#1}}%
   {\XXint\scriptscriptstyle\scriptscriptstyle{#1}}%
   \!\int}
\def\XXint#1#2#3{{\setbox0=\hbox{$#1{#2#3}{\int}$}
     \vcenter{\hbox{$#2#3$}}\kern-.5\wd0}}
\def\ddashint{\Xint=}
\def\dashint{\Xint-}

Throughout this section, we assume that the domain is an annulus. Once the existence of a neutral mode is established, we can then show that unstable modes appear when 
$k$ is slightly decreased.
Let us fix $N$ and vary $\lambda = -k^2$ in (\ref{Geq}).
Both $G$ and $c$ depend on $\lambda$, and their evaluation at the neutral parameter $\lambda=\lambda_0$ will be denoted by the subscript $0$.
Evaluation of (\ref{Geq}) at this parameter yields $\mathcal{L} G_0 = 0$, where $\mathcal{L}$ denotes the operator on the left-hand side of (\ref{SLSL}) with $\lambda=\lambda_0$.
By the properties of regular Sturm–Liouville problems, $G_0$ is real-valued.

The eigenvalue $c$ of (\ref{Geq}) can be expanded in a Taylor series. Thus if the first-order coefficient of the expansion
\begin{eqnarray}
c_i(\lambda)=\left. \frac{dc_i}{d\lambda}\right |_0(\lambda-\lambda_0)+O((\lambda-\lambda_0)^2)
\end{eqnarray}
is non-zero, this implies the emergence of an unstable mode. The eigenvalue $c$ behaves well in the complex plane except along the real axis since the two linearly independent solutions themselves remain well behaved; see \cite{Lin2003} for a mathematically rigorous discussion.

To proceed, we first differentiate (\ref{Geq}) with respect to $\lambda$ to obtain
\begin{eqnarray}
(\frac{r}{N^2+r^2}(rG_{\lambda})')'+(\lambda +\frac{rQ'}{U-c})G_{\lambda}+(1 +(\frac{rQ'}{U-c})_{\lambda})G=0.
\end{eqnarray}
Here, the subscript $\lambda$ denotes partial differentiation.
Evaluating this equation at $\lambda=\lambda_0$,
\begin{eqnarray}\label{Lg2}
\mathcal{L}G_{\lambda}|_0+\left. \left (1 +\left (\frac{rQ'}{U-c}\right )_{\lambda}\right |_0 \right )G_0=0.
\end{eqnarray}
Multiplying (\ref{Lg2}) by $rG_0$, subtracting $rG_{\lambda}|_0\mathcal{L}G_0=0$, and integrating over the domain, we obtain
\begin{eqnarray}
0=\int_{\Omega}(G_0\mathcal{L}G_{\lambda}|_0-G_{\lambda}|_0\mathcal{L}G_0)rdr+\int_{\Omega}\left. \left (1 +\left (\frac{rQ'}{U-c}\right )_{\lambda}\right |_0 \right)G_0^2rdr.
\end{eqnarray}
The first integral vanishes after integration by parts, so we get
\begin{eqnarray}\label{g0g0Kdc}
0=\int_{\Omega}G_0^2rdr+\left. K\frac{dc}{d\lambda} \right |_0,
\end{eqnarray}
where
\begin{eqnarray}\label{KrKi}
K=K_r+iK_i=\lim_{c_i\rightarrow 0^+}\int_{\Omega}\frac{rQ'G_0^2}{(U-\alpha-ic_i)^2}rdr\nonumber \\
=\dashint_{\Omega} \frac{rQ'G_0^2}{(U-\alpha)^2}rdr+i\pi \left. \left (\frac{rW_{\alpha,N}G_0^2}{|U'|} \right) \right|_{r=r_c}.
\end{eqnarray}
The dashed integral is the Cauchy principle integral. 
A heuristic derivation of (\ref{KrKi}) is well known in the shear flow instability community \cite[see][for example]{Drazin_Reid1981}. A rigorous mathematical justification, along the lines of \cite{Kumar2025}, can be straightforwardly done using the Plemelj-Sochocki formula.

Combining (\ref{g0g0Kdc}) and (\ref{KrKi}), we obtain
\begin{eqnarray}
\left. \frac{dc_i}{dk}\right |_0
=
-2\sqrt{-\lambda_0}\left. \frac{dc_i}{d\lambda}\right |_0
=-\frac{2\sqrt{-\lambda_0}K_i}{K_r^2+K_i^2}\int_{\Omega}G_0^2rdr.
\label{eq:dcidk}
\end{eqnarray}
A slight manipulation yields (\ref{dcidk}). 
As remarked earlier, $G_0$ has no zeros in the domain, so $\int_{\Omega}G_0^2rdr>0$.
Furthermore, from the assumptions,  $W_{\alpha,N}|_{r=r_c}=\frac{r_cQ''(r_c)}{U'(r_c)}$ is non-zero. In fact, if $W_{\alpha,N}$ surpasses the hurdle, then $W_{\alpha,N}|_{r=r_c}$ is strictly positive, and so is $\left. \frac{dc_i}{d\lambda}\right |_0$. Therefore, if $k$ is slightly decreased from its neutral value, $c_i$ must increase, as observed in figures \ref{fig:fig5}-(a) and \ref{fig:fig10}-(a).

\subsection{Proof of Theorem 2: trial functions}\label{sec:5.4}

To guarantee the existence of a physically relevant neutral solution, let us first consider how small $\lambda_0$ in (\ref{RayQ}) needs to be. The discussion differs between axisymmetric and non-axisymmetric perturbations.

The axisymmetric case is very similar to the corresponding analysis for parallel flows; a neutral mode exists if there is a trial function $\psi \in H^1_0$ such that 
\begin{eqnarray}\label{axiRRR}
R_{\alpha,0}(\psi)<0. 
\end{eqnarray}
If this condition holds, we have $\lambda_0 \leq R_{\alpha,0}(\psi)<0$, allowing us to set $k=\sqrt{-\lambda_0}$. 

For the non-axisymmetric case, a stronger condition is required: a neutral mode exists if there is a trial function $\psi \in H^1_0$ such that 
\begin{eqnarray}\label{nonaxiRRR}
R_{\alpha,N}(\psi)<-N^{-2}
\end{eqnarray}
for some $N$, say $N=N_+$. This condition guarantees $\lambda_0< -N^{-2}$ at $N=N_+$, so setting $k=\sqrt{-\lambda_0}$ yields a neutral solution with $N^{-1}<k$. Note that this neutral solution may not be the one we want, since physically $n=Nk$ must be an integer. Nevertheless, condition (\ref{nonaxiRRR}) is sufficient to ensure the existence of a physically relevant neutral solution. The reason is as follows. 
We first note that 
the eigenvalue $\lambda_0$ of the regular Sturm–Liouville problem depends continuously on $N$. 
Thus, if for some $N=N_-$ we have $\lambda_0> -N^2$, then there is an $N \in (N_-,N_+)$ such that $\lambda_0=-N^{-2}$ from the mean value theorem. 
We can indeed find such $N_-$ at sufficiently small $N$, since $-N^{-2}\rightarrow -\infty$ as $N\rightarrow 0$, while $\lambda_0$ remains finite. By taking $k=\sqrt{-\lambda_0}$ at the value of $N$ where $\lambda_0=-N^{-2}$ is satisfied, we obtain a neutral mode with $n=1$.

We now turn our attention to showing that the conditions in the theorem imply the existence of a neutral mode.
We first prove the annular case.
By assumption there is a constant $h\in \RealN$ such that $W_{\alpha,N}> h$ for all $r \in \Omega$. Then using (\ref{Rayq}) we can show that
\begin{eqnarray}
R_{\alpha,N}(\psi)<\frac{\int_{\Omega} \frac{r}{N^2+ r^2}|(r\psi)'|^2dr}{\int_{\Omega} \frac{1}{r}|r\psi|^2 dr}-h\leq \frac{\frac{1}{N^2+ r_i^2}\int_{\Omega} r|(r\psi)'|^2dr}{\int_{\Omega} \frac{1}{r}|r\psi|^2 dr}-h
\end{eqnarray}
The right hand side becomes $\frac{(\pi/\ln\eta)^2}{N^2+ r_i^2}-h$ upon substituting the trial function $\psi(r)=r^{-1}\sin\left (\pi\frac{\ln (r_i/r)}{\ln \eta} \right )$. 
This constant equals $-N^2$ when $h$ is set as (\ref{hthm2}), in which case condition (\ref{nonaxiRRR}) is satisfied.

The proof of the pipe case is similar, but employs a different trial function. The assumption
$W_{\alpha,N}>Cr^{p}$ for all $r\in \Omega$ implies
\begin{eqnarray}\label{RRRpipe}
R_{\alpha,N}(\psi)<\frac{\int_{\Omega}\{ \frac{r}{N^2+ r^2}|(r\psi)'|^2 -Cr^p\frac{1}{r}|r\psi|^2 \} dr}{\int_{\Omega} \frac{1}{r}|r\psi|^2 dr}.
\end{eqnarray}
With the choice of trial function $\psi=(1-r^2)$, which is suitable for $n=1$ modes, the right hand side of (\ref{RRRpipe}) evaluates to $6(\rho_N-h\rho_p)$.
Substituting $C$ in (\ref{hthm3}), this constant becomes $-N^{-2}$, which shows that (\ref{nonaxiRRR}) is satisfied.

\section{Conclusion and discussion} \label{sec:6}

We studied the inviscid stability problem (\ref{Geq}) for axisymmetric base flows in annuli and pipes.
The main results are the two theorems presented in section~\ref{sec:3}, which provide simple analytical tools to estimate the behaviour of the eigenvalue $c$. By using those theorems complementarily, we can estimate the location of the neutral point in the parameter space, as demonstrated through the analysis of model flows.

Theorem \ref{KA} provides a sufficient condition for stability. The KA-II condition further extends the range of parameters for which stability can be guaranteed, beyond the classical KA-I result of \cite{Batchelor_Gill_1962}.
As shown in section~\ref{sec:5.1}, the KA-II condition follows from elementary algebraic manipulations together with the application of Poincar\'e-type inequality. In practice, the condition can be assessed by checking whether the function $W_{\alpha,N}$ defined in (\ref{defW}) remains below a threshold
$H$ (see figures~\ref{fig:fig4}-(b) and \ref{fig:fig9}-(a)). The form of $H$ differs between annular and pipe flows and also depends on whether the disturbance is axisymmetric or non-axisymmetric.

Theorem~\ref{Hannulus} gives a sufficient condition for instability. For annular flows, a constant hurdle $h$ can be introduced, as in \cite{Deguchi_Hirota_Dowling_2024} for parallel flows. Surpassing the hurdle $h$ by $W_{\alpha,N}$ implies instability (figure~\ref{fig:fig4}-(a)). 
However, for pipe flows, the constant hurdle is ineffective as $W_{\alpha,N}$ behaves like $r^2$ near the origin. Accordingly, we constructed the instability condition so that $h$ is modified to take a power-law dependence on $r$ (figure \ref{fig:fig9}-(c)). As shown in the proof in section~\ref{sec:5.2}, the specific form of $h$ depends on the choice of trial function. While various forms of $h$ are possible, our choice offers an optimal balance between simplicity and effectiveness.

In step (i) of the proof of Theorem \ref{Hannulus} (section~\ref{sec:5.2}), Sturm–Liouville theory plays a key role in establishing that a neutral solution exists when the minimum of the Rayleigh quotient (\ref{RayQ}) is sufficiently small. In step (ii) (section~\ref{sec:5.3}), a perturbation analysis is then used to estimate the change in the growth rate caused by a slight variation in the wavenumber and to identify the onset of instability (the red dashed lines in figures \ref{fig:fig5}-(a) and \ref{fig:fig10}-(a)). 
In section~\ref{sec:5.4}, test functions are selected to satisfy (\ref{axiRRR}) for the axisymmetric case and (\ref{nonaxiRRR}) for the non-axisymmetric case.

The theoretical results have been applied to the annular model flow (section \ref{sec:4.1}) and the pipe model flow (section \ref{sec:4.2}). In both cases, the stable and unstable regions in parameter space predicted by Theorems \ref{KA} and \ref{Hannulus}, respectively, are consistent with the exact stability analyses obtained from the eigenvalue problem (\ref{Geq}).
Furthermore, these results agree with stability analyses based on the linearised Navier-Stokes equations in the limit of sufficiently large Reynolds numbers. 
The theorems require only a simple graphical analysis, which provides a rough estimate of stability in parameter space and 
thereby reduces the number of eigenvalue computations needed. They are not limited to model flows and could be applied in a variety of situations. 
For example, they can be used for local temporal stability analyses of slowly spatially developing flows, such as diverging pipe flows \citep{Sahu_2005}.

It should be noted that the results of the theorems are valid only for stability analyses based on the inviscid approximation. As exemplified by Tollmien–Schlichting waves, viscosity can contribute to instability, so even within the stable regions predicted by Theorem~\ref{KA}, the full stability problem without approximation may exhibit instability. One approach to obtain an analytic estimate of instabilities that persist in the presence of viscosity is to use a Wentzel–Kramers–Brillouin (WKB) type approximation \citep{Kirillov_2014,Kirillov_Mutabazi_2025}. 
However, this method is valid only for short wavelengths, where shear flows are typically stable, and hence does not capture the instabilities predicted by Theorem \ref{Hannulus}.

Finally, we discuss the stability of axisymmetric jets, which are not covered by our theorems but were the main subject of the study by \cite{Batchelor_Gill_1962}. Unbounded domains like jets correspond to a singular Sturm–Liouville problem, which is likely to be more difficult than the pipe case and would require additional steps in the proof. Even ignoring this mathematical technicality, there are more fundamental difficulties.
First, Poincar\'e-type inequalities cannot be applied in unbounded domains. Therefore, in Theorem \ref{KA}, it is difficult to improve the stability results.
Also, the form of $h$ given in Theorem~\ref{Hannulus} is not optimal for detecting unstable modes, as $W_{\alpha,N}$ tends to zero in the far field. This is illustrated, for example, from figure 3 of \cite{Batchelor_Gill_1962}, where $N^2$ corresponds to  $W_{\alpha,N}$ in the Schlichting jet
\begin{eqnarray}\label{Sjet}
U(r)=\frac{1}{(1+r^2)^2}.
\end{eqnarray}

That said, there is a convenient way to design suitable trial functions for the jet flows, as shown in Appendix \ref{sec:AppB}. For the Schlichting jet, the integral in condition (\ref{nonaxiRRR}) can be calculated analytically, showing that instability occurs, which is consistent with the Navier-Stokes computations of \cite{Lessen_Singh_1973}.
However, in general, numerical integration is required to use (\ref{nonaxiRRR}). 
Whether simple conditions like those we derived for annular and pipe flows can be obtained for jets remains a topic for future work. If successful, this could facilitate the stability analysis of more practically relevant flows, such as plumes \citep{Chakravarthy_Lesshafft_Huerre_2015}.

\backsection[Acknowledgements]{
This research was supported by the Australian Research Council Discovery Project DP230102188. }

\backsection[Declaration of Interests]{
The authors report no conflict of interest.
}

\appendix
\section{Poincar\'e-type inequalities}\label{sec:AppA}

Poincar\'e-type inequalities play a crucial role in the derivation of our theorem. Although their derivation is elementary, we present it here for completeness. Moreover, these inequalities are useful for providing a physical interpretation of inviscid stability in terms of the reciprocal Rossby-Mach number, which is discussed in section \ref{sec:AppA4}.

\subsection{Derivation of (\ref{fineq})}\label{sec:AppA1}
The most straightforward approach to find the best constant in the inequality (\ref{fineq}) is to solve the eigenvalue problem
\begin{eqnarray}
(\frac{r}{N^2+ r^2}(rf)' )'+\lambda f=0,\label{eigenlambda}
\end{eqnarray}
with boundary conditions $f(r_i)=f(r_o)=0$. Since this is a regular Sturm–Liouville problem, the eigenfunctions form a complete orthonormal basis for expanding $f \in H_0^1$. Consequently, the inequality (\ref{fineq}) holds with $\kappa_N^2$ replaced by the smallest eigenvalue $\lambda$ of (\ref{eigenlambda}).  Unfortunately, this eigenvalue problem cannot be solved using elementary functions.
Thus, we estimate the constant from below: 
\begin{eqnarray}
\int^{r_o}_{r_i} \frac{r}{N^2+ r^2}|(rf)'|^2 dr
\geq
\frac{1}{N^2+ r_o^2}\int^{r_o}_{r_i} r|(rf)'|^2 dr\nonumber \\
\geq \frac{(\pi/\ln \eta)^2}{N^2+ r_o^2}\int^{r_o}_{r_i} r^{-1}|rf|^2dr.
\end{eqnarray}
The second inequality follows from
\begin{eqnarray}
\int^{r_o}_{r_i} r|(rf)'|^2 dr
\geq \lambda \int^{r_o}_{r_i} r^{-1}|rf|^2dr.\label{ineq33}
\end{eqnarray}
Here, $\lambda=(\pi/\ln \eta)^2$ is the minimum eigenvalue of the Sturm–Liouville problem $(r(rf)')'+\lambda f=0$,  $f(r_i)=f(r_o)=0$. The corresponding eigenfunction is $f(r)=r^{-1}\sin\left (\pi\frac{\ln (r_i/r)}{\ln \eta} \right )$. 

\subsection{Derivation of (\ref{ineq1})}\label{sec:AppA2}
The best constant in the inequality 
\begin{eqnarray}
\int_{0}^{1} \frac{1}{r}|(rf)'|^2 dr\geq \lambda \int_{0}^{1} \frac{1}{r}|rf|^2dr
\end{eqnarray}
can be found by the associated Sturm–Liouville eigenvalue problem $(r^{-1}(rf)')'+\lambda f=0$. The solution of this differential equation that remains regular at $r=0$ is $f(r)=J_1(\sqrt{\lambda}r)$, where $J_1$ is the Bessel function of the first kind of order one. Applying the boundary condition $f(1)=0$, the minimum eigenvalue is found to be $\lambda=j_{1,1}^2$. 
Note that from the regularity condition (\ref{taylor}), the integral appearing on the left-hand side of the inequality is finite.

\subsection{Derivation of (\ref{ineq2})}\label{sec:AppA3}
We can find (\ref{ineq2}) by showing 
the best constant in the inequality
\begin{eqnarray}\label{ineqjn1}
\int^1_0 \{r|(rf)'|^2 +\frac{n^2}{r}|rf|^2\}dr\geq \lambda \int^1_0 r|rf|^2dr.
\end{eqnarray}
is $\lambda=j_{n,1}^2$. Here, $j_{n,1}$ denotes the first positive zero of the Bessel function of the first kind of order $n$, and $j_{n,1}> j_{1,1}$ for $n\geq 2$.

The Sturm–Liouville eigenvalue problem associated with (\ref{ineqjn1}) is $(r(rf)')'-n^2f+\lambda r^2f=0$, which admits a regular solution $f(r)=r^{-1}J_n(\sqrt{\lambda}r)$. The boundary condition $f(1)=0$ is satisfied when $\lambda=j_{n,1}^2$.

\subsection{Comments on the reciprocal Rossby-Mach number}\label{sec:AppA4}

\cite{Stamp_Dowling_1993} proposed that inviscid stability theory is effective for the physical interpretation of observational data of Jupiter’s atmosphere. Later, \cite{Dowling2020} defined the reciprocal Rossby-Mach number $M^{-1}$ such that neutrality occurs when $M^{-1}=1$ in cases where the KA stability conditions sharply discriminate stability. Jupiter’s atmosphere appears to realise the condition $M^{-1}=1$. However, 
the missing piece at that stage was to establish a criterion that detects instability and to verify under what conditions $M^{-1}=1$ indeed corresponds to neutrality; this was achieved by \cite{Deguchi_Hirota_Dowling_2024}. For a recent review of the historical development of this problem, see \cite{RD26}.

Although the reciprocal Rossby-Mach number can also be defined for the problems considered in this paper, its introduction requires some care, and we therefore chose not to include it in the main text. 
In particular, the need to distinguish between annular and pipe geometries, as well as between axisymmetric and non-axisymmetric perturbations, complicates the discussion; therefore, we restrict our attention here to the annular case assuming non-zero fixed $N$ (so the perturbation is non-axisymmetric).

Let $\kappa_N^2$ denote the smallest  eigenvalue $\lambda$ of (\ref{eigenlambda}),  and let  $\hat{f}$ be the associated eigenfunction. We then define the reciprocal Rossby-Mach number by
\begin{eqnarray}\label{defM}
M^{-1}=\frac{W_{\alpha,N}}{\kappa_N^2+N^{-2}}
=\frac{rQ'/(\kappa_N^2+N^{-2})}{U-\alpha}.
\end{eqnarray}
Recalling that $\alpha$ typically corresponds to the drift speed of the neutral wave, the denominator of the right hand side of (\ref{defM}) represents the flow speed relative to this drift speed. Hence, viewing the numerator as a `sound speed' leads to an analogy with gas dynamics. In dimensional form, it indeed has the dimensions of velocity, noting that both $N$ and $1/\kappa_N$ have dimensions of length. Physically, the factor $1/\sqrt{\kappa_N^2+N^{-2}}$ can be interpreted as an estimate of the maximum size of a helical perturbation with a given pitch.

The flow is stable if $0\leq M^{-1}\leq 1$ everywhere in the domain (i.e. it is supersonic).
Indeed, under this assumption, equation (\ref{Zeq}) yields
\begin{eqnarray}
0=\int_{\Omega} \{\frac{r}{N^2+ r^2}|(rG)'|^2 +\frac{k^2}{r}|rG|^2-(\kappa_N^2+N^{-2})M^{-1}Z\frac{|rG|^2}{r} \}dr\nonumber \\
>(k^2-N^{-2})\int_{\Omega} \frac{|rG|^2}{r} dr=k^2(1-n^{-2})\int_{\Omega} \frac{|rG|^2}{r} dr,
\end{eqnarray}
which cannot be satisfied by non-axisymmetric unstable perturbations.
Note that, in the second line, we used the inequality (\ref{fineq}), which follows from the Sturm-Liouville problem (\ref{eigenlambda}).

We can also show that the flow is unstable if $M^{-1}>1$ everywhere in the domain (i.e. it is subsonic).
Substituting $\hat{f}$ into the Rayleigh quotient defined in (\ref{Rayq}),
\begin{eqnarray}\label{A8A8}
R_{\alpha,N}(\hat{f})=\frac{\int_{\Omega}\{ \frac{r}{N^2+ r^2}|(r\hat{f})'|^2 -(\kappa_N^2+N^{-2})M^{-1}\frac{1}{r}|r\hat{f}|^2 \} dr}{\int_{\Omega} \frac{1}{r}|r\hat{f}|^2 dr}\nonumber\\
<\frac{\int_{\Omega}\{ \frac{r}{N^2+ r^2}|(r\hat{f})'|^2 -(\kappa_N^2+N^{-2})\frac{1}{r}|r\hat{f}|^2 \} dr}{\int_{\Omega} \frac{1}{r}|r\hat{f}|^2 dr}=N^{-2}.
\end{eqnarray}
Comparing the above with the condition (\ref{nonaxiRRR}) establishes the existence of an unstable mode.
Here, to derive the second line of (\ref{A8A8}), we used the fact that $\hat{f}$ attains equality in the inequality (\ref{fineq}).

From the above considerations, it follows that if the base flow is designed so that $M^{-1}$ is constant, then $M^{-1}=1$ (i.e. the sonic case) constitutes the stability boundary. Note, however, that this conclusion is restricted to fixed $N$. Physically relevant stability analysis requires consideration of different $N$ and axisymmetric perturbations, and there appears to be no simple criterion for identifying which type of perturbations is the most dangerous.

We furthermore remark that $\kappa_N$ used in this section does not admit simple closed form expression. As discussed in section~\ref{sec:AppA1}, the analytic formula (\ref{kappaN}) used in the main text only represents a lower bound estimate of the best constant adopted here. Consequently, plotting $M^{-1}$ requires solving the eigenvalue problem (\ref{eigenlambda}) numerically, which complicates the graphical analysis in section \ref{sec:4}.

\section{Analysis of the Schlichting jet}\label{sec:AppB}

In the limit of large $N$,  condition (\ref{nonaxiRRR}) reduces to the following.
\begin{eqnarray}
\frac{\int_{\Omega} \{r|(r\psi)'|^2+\frac{(rU')'}{U-\alpha}|r\psi|^2 \}dr}{\int_{\Omega} \frac{1}{r}|r\psi|^2dr}<-1.
\end{eqnarray}
Choosing the trial function $\psi=U$, 
the above inequality can be transformed into
\begin{eqnarray}\label{rUrU}
\int_{\Omega} \{r((rU)')^2+\frac{(rU')'}{U-\alpha}(rU)^2  +rU^2\}dr<0.
\end{eqnarray}
For the Schlichting jet (\ref{Sjet}), we obtain $\alpha=4/9$ and $\frac{r}{U-\alpha}(rU')'
=\frac{
-72r^2
}{(r^2+1)^2(2r^2+5)}$. Using those results, the left hand side of (\ref{rUrU}) can be evaluated as
\begin{eqnarray}
-\int^{\infty}_0 \frac{8(2r^2-1)}{(r^2+1)^5(2r^2+5)}r^3 dr=-\text{ln}\left (\frac{5^3 2^{\frac{8}{81}}10^{\frac{77}{81}}}{32}\right )+\frac{97}{27}\approx-0.0273.
\end{eqnarray}

\bibliographystyle{jfm}  
\bibliography{Reference}  

\end{document}